\documentclass[a4paper,10pt]{article}
\usepackage{cite}
\usepackage{a4}
\usepackage{ulem}
\usepackage{color}
\usepackage{epsfig}
\usepackage{graphicx}
\usepackage{subfigure}
\usepackage{ulem}
\usepackage{amsfonts}
\usepackage{epsfig,bm}
\usepackage{graphicx}
\usepackage{comment}
\begin{document}
\title{Geometric Perspective of Entropy Function:\\ Embedding, Spectrum and Convexity\footnote{This 
note is a preliminary version of the talk given on 30 March 2005 as High Energy Theory Group Seminar 
at the Department of Physics, Indian Institute of Technology Kanpur, India.}}
\author{}
\vspace{0.10 cm}
\date{Bhupendra Nath Tiwari\thanks{\noindent E-mail: bntiwari.iitk@gmail.com}\\
\vspace{0.50 cm}
INFN-Laboratori Nazionali di Frascati\\
Via E. Fermi 40, 00044 Frascati, Italy.}
\vspace{0.5cm}
\maketitle
\begin{abstract}
From the perspective of Sen entropy function, we study the
geometric and algebraic properties of a class of (extremal) black
holes in $ D \geq 4 $ spacetimes. For a given moduli space manifold, 
we describe the thermodynamic geometry away from attractor fixed 
point configurations with and without higher derivative corrections.
From the notion of embedding theory, the present investigation
offers a set of generalized complex structures and associated
properties of differentiable manifolds. We have shown that the
convexity of arbitrary entropy function can be realized in an
extended subfield of the eigenvalues of the Hessian $\mathcal B$ 
of Sen entropy function. Thus, the spectra of $\mathcal B$ are
analyzed by defining Krull of the corresponding semisymplectic
algebras. From the framework of commutative algebra, we find that
the convex hull of the eigenvalues defines a generalized spectrum 
of $\mathcal B$. The corresponding complexification is established
for finitely many eigenvalues of $ \mathcal B $. For the minimally
extended subfield, we show that the spectrum of $ \mathcal B $
reduces to the thermodynamic type spectra, at the attractor fixed
point(s). In the limit of $ AdS_2 \times S^{D-2} $ near horizon 
geometry, the attractor flow analysis offers the stability of 
arbitrary extremal black hole. From the perspective of string 
compactifications, our investigation implies a set of deformed 
S-duality transformations, which contain both the duality invariant 
charges and monodromy invariant parameters. The role of the algebraic
geometry is discussed towards the viewpoints of attractor stability 
conditions, rational conformal field theory, elliptic curves, deformed 
quantization(s), moduli manifolds and Calabi-Yau.
\end{abstract}

\vspace{1.50cm}

\textit{\textbf{Keywords}: Entropy Function; Embedding; Spectrum;
Convexity; Calabi-Yau; Attractor Stability; S-duality; Deformed
Quantization; Generalized  Geometry; Higher Derivative Gravity;
Black Hole Physics.} 
%
\newpage
\begin{Large} \textbf{Contents:} \end{Large}

\begin{enumerate}
\item{Introduction.}
\item{Real and Complex Geometry.}
\item{Sen Entropy Function.}
\item{Embedding and Entropy Function.}
\item{Spectrum and Convexity.}
\item{Calabi-Yau Geometry.}
\item{Conclusion and Outlook.}
\end{enumerate}
\section{Introduction}\label{one}
Definition of the surface gravity follows from the existence of 
a limiting force, which must be exerted at infinity, in order to
hold a unit test mass in place, when approaching the horizon of
a static black hole. The surface gravity remains a constant over
an imaginary surface inside which there cannot exist an object
that can ever escape to the outside world. Such an imaginary
surface defines the event horizon of a black hole
\cite{HawkingWald,HawkingWald1}. Mathematically, it has been shown
that the surface gravity remains constant over the horizon, for a
diffeomorphism covariant Lagrangian \cite{Wald,Wald1,Wald2}.
Further, for any time $ t $ in classical general relativity, the
change in the horizon area remains a positive definite quantity \cite{Hawking}.
For a given black hole, there exists an equivalence of the energy
to the mass, entropy to the horizon area, volume to the angular
momentum and the number of moles to the charges. This amounts to
the statement that black holes are thermodynamical objects
\cite{Bekenstein,Bekenstein1}. In particular, we wish to exploit
the fact that the black hole entropy $ S_{BH} $ is a
thermodynamical quantity. Recently, Sen has obtained a procedure
to compute the entropy of a class of extremal black holes, from
the entropy function method \cite{SenEntropyFunction,
SenEntropyFunction1,SenEntropyFunction2,SenEntropyFunction3,
SenEntropyFunction4,SenEntropyFunction5,SenEntropyFunction6,
SenEntropyFunction7,SenEntropyFunction8}. In this direction, our 
study is extensively applicable to all black branes, as long as 
the entropy function is known. Moreover, an extension of the entropy
function formalism has been made for $ D_1D_5 $ and $ D_2D_6NS_5 $
non-extremal configurations \cite{GarousiGhodsiCai,
GarousiGhodsiCai1,GarousiGhodsiCai2}. In the throat approximation,
these solutions respectively correspond to the Schwarzschild black
holes in $ AdS_3 \times S^3 \times T^4 $ and $ AdS_3 \times S^2
\times S^1 \times T^4 $.

The entropy function method has emerged as one of the most
powerful technique for calculating the entropy of a large class of
extremal black holes in $ D \geq 4 $. Interestingly, such extremal
black holes possess $ AdS_2 \times S^{D-2} $ near horizon
geometry. It has been known \cite{SenEntropyFunction,Sen0} that
the entropy function generalizes the standard attractor mechanism 
\cite{FerraraStrominger1,FerraraStrominger2,FerraraStrominger,FerraraStrominger3}, 
\textit{viz.} the horizon data depend only on the charges of the black hole,
and remain independent of the corresponding asymptotic values of
the scalar fields. Subsequently, the entropy function method
provides an interesting front for incorporating various higher
derivative corrections. Moreover, such a consideration takes an
account of the corrections to the holomorphic prepotential, and
thereby it leads to the theory of generalized prepotential $
\mathcal F $ \cite{CardosoWit}. As examined in section $4$, the
reason follows from the fact that Legendre transformation of the
black hole entropy $ S_{BH} $, when performed with respect to the
radial electric fields $ \lbrace e_i \rbrace $, leads to the
definition of the generalized prepotential. It is worth mentioning
that $ S_{BH} $ has found further importance from the perspective
of the topological string partition function
\cite{VGVD,VGVD1,VGVD2,VGVD3}. On the other hand, the charged
black holes in string theory possess the statistical entropy
$S_{BH} = \ln \Omega(\overrightarrow{p}, \overrightarrow{q}) $,
where $ \Omega(\overrightarrow{p}, \overrightarrow{q})$ is the
degeneracy of the elementary string states
\cite{SenDabholkar,SenDabholkar1, SenDabholkar2,SenDabholkar3}.
For the small black holes, it is known \cite{SenPeet,SenPeet1}
that the leading order entropy vanishes identically and one has
$S_{BH} = 0$, in the supergravity limit. Most of these
formulations involve a heavy use of the supersymmetry algebra in
order to analyze the properties of the underlying higher order
$\alpha^{\prime}$-corrections. In contrast to the above, for a
given Lagrangian density, Sen entropy function method solely
depends on the near horizon field configuration and thus the bosonic
content of the theory enables one to evaluate the entropy of an extremal black hole. 
Interestingly, a class of $\alpha^{\prime}$-corrections to the black hole
entropy, arising from an underlying string theory compactification, 
agrees with the corresponding entropy obtained from the topological 
string partition function consideration \cite{FerraraStrominger3,VafaAntoniadisNarain1}.

For a given black hole configuration, such $ \alpha^{\prime} $-corrections 
are governed by the generalized prepotential $ \mathcal F $
\cite{CardosowdeWit,CardosowdeWit1,CardosowdeWit2,CardosowdeWit3,
CardosowdeWit4}. In this sense, the higher derivative terms play a
special role from the perspective of effective string actions.
In order to produce the required black hole entropy $S_{BH}$, the
formulation of Sen entropy function not only offers a few terms
which appear in the effective Lagrangian density, but it also examines
the role of the complete set of higher derivative terms, which
could affect the near horizon geometry of an extremal black hole
\cite{Sen3}. Furthermore, it is known that the entropy function
method is an analysis of the equations of motion, which does not
make a direct use of supersymmetry. Therefore, the evaluation of
the entropy only requires to know certain specific structures of
the higher derivative terms, which depend on the compactifying
internal space. Importantly, the entropy function method remains
duality invariant, as the equation of motions are so, thus it provides 
a duality invariant formula for Wald entropy $S_{BH}$ of the chosen 
black holes \cite{Wald,Wald1,Wald2}. It is worth mentioning that the 
entropy function method is applicable only to those Lagrangian density 
$ \mathcal L $, which can be expressed in terms of gauge invariant
field strengths and does not explicitly involve the gauge fields of
the theory. For example, such a situation arises in the presence of
Chern Simon corrections to the Lagrangian density. In such a case, 
one can try to (i) remove the non-covariant terms, for instance
by going to the dual field variables and thereby (ii) bring out the
Lagrangian density $ \mathcal L $ into the required form. Interestingly, 
one of such an example is the BTZ black hole under the Chern Simons and 
higher derivative $\alpha^{\prime}$-corrections. In this case,
it is known \cite{Sen-Sahoo2} that the gauge field $ \lbrace A_i
\rbrace $ are expressible in terms of the field strength tensor $
F_{\mu \nu}$. Following Wald entropy formula
\cite{Wald,Wald1,Wald2}, if it is not possible to eliminate an
explicit gauge field dependence from the Lagrangian density, then
the entropy function method needs a further generalization.

From the perspective of the attractor stability of $ AdS_2 \times
S^{D-2} $ near horizon geometry, it is worth analyzing the physical 
properties of higher derivative gravity. In this investigation, 
we consider an arbitrary theory of the higher derivative gravity, 
which is coupled with the following sets of fields and their 
respective attractor fixed point horizon values: (i) scalar fields $ \lbrace
\phi_s \rbrace \rightarrow \lbrace u_s \rbrace $, (ii) electric-
magnetic field tensor $ F_{\mu \nu}^i $ with $ \lbrace F_{rt}^i
\rightarrow e_i , F_{\theta \phi}^i \rightarrow \frac{p_i}{4
\pi}sin \theta  \rbrace $ and (iii) $ AdS_2 \times S^{D-2} $
parameters $ \lbrace v_i \rbrace \rightarrow \lbrace v_i^0 \rbrace
$. For given horizon data $\{u_s, e_i,p_i,v_i\}$ of an extremal
black hole, let $ F(\overrightarrow{u},\overrightarrow{v},
\overrightarrow{e},\overrightarrow{p}) $ be Sen entropy function, 
then the corresponding attractor entropy $ S_{BH}(\overrightarrow{q},
\overrightarrow{p}) $ can be computed by requiring the existence 
of asymptotically flat extremal black hole solution. See section 
$3$ for further details. One of the motivational question which we 
wish to address in this work is the following: Is $ AdS_2 \times S^{D-2} $ 
near horizon geometry a stable attractor? Namely, we examine whether 
there really exist an interpolating solution between the flat asymptotic 
geometry and $ AdS_2 \times S^{D-2} $ near horizon geometry. The answer is 
known in certain two derivative theories, see for instance 
\cite{SenEntropyFunction3,SenEntropyFunction6}. In order to have
an interpolating solution between the flat asymptotic geometry and
$ AdS_2 \times S^{D-2} $ near horizon geometry, the Hessian matrix
of the entropy function must be positive definite metric at the
critical point(s). Further, it is natural to ask: Does there exist
certain generalizations in order to incorporate higher derivative
corrected action(s)? As the next step, in order to answer such a
question, we offer the algebraic geometric perspective in section $5$.

In this short note, we set out to explore the attractor stability of
various extremal black holes with a given set of $ \alpha^{\prime}
$-corrections. For given entropy function, we define moduli dependent 
thermodynamic geometry, which as an intrinsic geometry leads to 
interesting extensions of the thermodynamic configuration, from 
the viewpoint of the attractor mechanism. In particular, such a 
geometric extension renders into a moduli dependent (intrinsic) 
manifold, which are known since the thermodynamic notion of 
$ \mathcal N = 2 $ supergravity theories \cite{FerraraStrominger1}. 
Further, the stability of the central charge $ Z $ and the 
corresponding covariant derivatives have been of a constant 
interest in various supergravity theories \cite{XX}. From the 
perspective of thermodynamic geometry \cite{Ruppenier,
Ruppenier1,Ruppenier2}, there exists an intrinsic Riemannian
metric structure, which could be defined as the negative Hessian
matrix of the attractor entropy. For a class of extremal and non-extremal 
black holes in string theory and M-theory \cite{bnt1,bnt2,bnt3,bnt4,bnt5,bnt6,bnt7}, 
it has been observed that the thermodynamic geometry offers a well-defined 
notion of statistical stability and thereby the role of higher 
derivative corrections at a given attractor. Indeed, the effect of $R^2$ 
and $R^4$ spacetime corrections are known for certain non-extremal black holes
in string theory, see for instance \cite{CardosowGhodsi, CardosowGhodsi1,
CardosowGhodsi2,CardosowGhodsi3,CardosowGhodsi6, CardosowdeWit3}.
From the viewpoint of the intrinsic algebraic geometry, we show that 
the framework of Sen entropy function method \cite{Sen4,Sen41} 
offers an efficient characterization of the stability under 
both the effects of finitely many (i) stringy 
$\alpha^{\prime}$-corrections and (ii) quantum loops.

In order to extend the role of the thermodynamic geometry away
from the attractor fixed point(s), we need to consider a set of
scalar fields $ \lbrace \varphi^a \rbrace $. As illustrated in
section $2$, the K\"{a}hler moduli can be explicitly characterized
by a set of complex holomorphic coordinates $ \lbrace z^i \rbrace
$ and thus the K\"{a}hler metric can be expressed as $ G_{ab} d
\varphi^a d \varphi^b= \frac{\partial^2 K}{\partial z^i \partial
\overline{z}^j} dz^i d \overline{z}^j $. Away from the attractor
fixed points \cite{Kallosh97}, there exists an embedding $ R^k
\hookrightarrow R^k \times \mathcal M_{\varphi} $. Geometrically,
such a configuration describes a manifold of dimension $ k= 1+ 2l
$, where $ l $ is the number of the electric-magnetic charges. For
given thermodynamic variables $ \lbrace q_A, p^A \rbrace $, the
asymptotic moduli fields $ \lbrace \varphi^a_{\infty} \rbrace $
modify the black holes thermodynamics as $ dM= T dS+ \psi^A dq_A+
\chi_A dp^A- \Sigma_a d \varphi^a_{\infty} $, where $
\varphi^a_{\infty} \in \mathcal M_{\{\varphi^a_{\infty}\}} $
\cite{Kallosh97}. For an arbitrary scalar moduli $ \mathcal
M_{\varphi} $, one needs to extend the partial derivative $
\partial_a $ to the K\"{a}hler covariant derivative $ \nabla_a $,
corresponding to the scalar fields with respect to the K\"{a}hler
metric $ G_{ab} $ on $ \mathcal M_{\varphi} $. In the limit when
temperature $ T \rightarrow 0 $, the underlying black hole becomes 
extremal and thus the components of Ruppenier metric $ S_{ab}
\rightarrow \infty $. Thus, it turns out that the standard notion
of black hole thermodynamics breaks down \cite{Kallosh97}. In
order to deal with black hole thermodynamics near the extreme,
we need to consider a renormalized notion \cite{Kallosh97} of 
Ruppenier metric $ S_{ab}= -\nabla_a \nabla_b A(p, q, \varphi) $
and Weinhold metric $ W_{ab}= \nabla_a \nabla_b M(p, q, \varphi) $. 
From the perspective of the moduli space geometry, the corresponding 
Weinhold geometry of an extremal black holes may be defined in terms 
of the black hole potential $ V_{BH} $. Following Ref.\cite{Kallosh97}, 
it turns out that Weinhold metric reduces to the following expression 
$ W_{ab}= \frac{1}{2 \sqrt{V_BH}}\nabla_a \nabla_b V_{BH}(p, q, \varphi) $.

One of the main interest of the present note is thus to explore
the stability of various higher derivative corrected black hole
configuration. From the perspective of the attractor stability, we
wish to expose the role of algebraic geometry, \textit{viz.} we shall
offer possible modification(s) of the thermodynamic geometry away
from the attractor fixed point(s). We take a rigorous account of
the scalar fields, which arise as an artifact of the underlying
string theory compactification. From the entropy function method,
Sen has demonstrated \cite{SenEntropyFunction} that the entropy is
the horizon value of the corresponding entropy function. Thus, the
central idea of the present work is kept to explain the following
question: What happens to the thermodynamic intrinsic geometry
\cite{Ruppenier,Ruppenier1,Ruppenier2}, when one moves away 
from the attractor fixed point(s)? In the context of 
$AdS_2 \times S^2 $ near horizon geometry, the effective 
potential $ V_{BH} $ can be further expressed in terms of the
central charge $ Z_{BH} $ of the extremal black hole
\cite{Hatta07}. Subsequently, for given K\"{a}hler potential $K$,
the covariant derivatives \cite{Hatta07} take the following form $
\nabla_a Z= (\partial_a+ \frac{1}{2} \partial_a K) Z $. As the
main interest of the present work is to examine the attractor
stability of an extremal black hole in arbitrary spacetime
dimension $D$, thus it suffices to consider an extreme K\"{a}hler
moduli such that the K\"{a}hler potential $K$ remains near the
chosen critical point(s), under the fluctuations of the moduli. 
Subsequently, the examination of the attractor stability of 
extremal black holes can be considered as the limiting map 
$ \nabla_a \rightarrow \partial_a $. Algebraically, we have 
shown that such a map involves the replacements $ \nabla_a
\longrightarrow \partial_a $, for a given set of scalars $ X_a =
\lbrace X_0, X_1,.., X_n \rbrace $. Consequently, there exists a
class of shifted S-duality transformations and thus such a
prolongment arises as the local transformations on $ {\cal M}:=
\mathcal M_\phi \otimes \mathcal M_2 \otimes \mathcal M_{2n} $.
See section $4$ for an explicit construction of $ {\cal M}$. 
Indeed, the section $4$ shows that the electric and magnetic 
frames are described by the maps $(0,p_i) \longleftarrow ( q_i,
p_i ) \longrightarrow (q_i, 0) $. This motivates to analyze the
question that, under what condition(s), the Hessian matrix
$\mathcal B$ of the entropy function of an extremal black hole
remains well-defined and locally convex. In the limit of the
minimal field of the eigenvalues of $\mathcal B$, such a
consideration offers both the geometric and algebraic properties,
pertaining to the attractor stability of arbitrary $ D $
dimensional (extremal) black holes with and without higher
curvature spacetime corrections.

For a given entropy function, $\mathcal B$ defines a
semisymplectic geometry, which exhibits the generalized atlas,
generalized Legendre transform, generalized symplectomorphism,
semi-product and strong stability. Such a geometry describes
thermodynamic geometry together with a non-trivial intertuning of
the moduli space geometry. Physically, the underlying black hole
configuration corresponds to a moduli dependent interacting
statistical system. As a set of embeddings and their well-defined
compositions, it reduces to the (generalized) symplectic
geometry, which possesses a set of generalized complex structures.
From the perspective of commutative algebra, we have shown that
the convexity of an extremal black hole entropy function can be
realized in the minimally extended (sub)field of the eigenvalues of $
\mathcal B $. For a given $\mathcal M$, it follows that the spectrum
of $\mathcal B$ can be analyzed by defining Krull of the underlying 
algebra of $ \mathcal B $. Thus, the analysis of such a generalized 
convexity and spectrum is interesting from the perspective of the 
commutative algebra. Further, we notice that the convex hull, 
and thus the underlying complexification of generalized spectrum 
can be defined in the minimally extended algebraic field, 
as an extended set of the eigenvalues of $ \mathcal B $. Finally, 
it is not difficult to show that the above spectra reduce to a 
thermodynamic like spectrum at the attractor fixed point(s). 
In this case, the underlying geometry renders to the standard 
Riemannian geometry at the extremum value of the entropy function. 
Furthermore, it is possible to demonstrate that both the above 
mentioned geometries have the same non-zero complexifiable 
joint spectrum. Herewith, we find an agreement with Sen entropy 
extrimization principle. Physically, such an analysis explains 
attractor stability of $ AdS_2 \times S^{D-2} $ near horizon 
geometry, pertaining to the chosen extremal black hole. 
Hereby, we offer an account of the stringy effects, which
arise via an introduction of the scalar fields, gauge fields and 
arbitrary covariant higher derivative terms, in the gravity side. 
Form the perspective of string theory, the respective structures are
exploited in the sections $4-6$. In the case of the Calabi-Yau (CY)
compactification, we have explicitly shown in section $6$ that
the present consideration implies a set of deformed S-duality
transformations, which contain both the gauge charges and
monodromy invariant parameters. In principle, this opens up a new
avenue towards the compatibility structures and deformed
quantizations on CY manifolds.

The rest of the presentation is organized as follows. In section $2$,
we offer a set of definitions pertaining to the basic notion of 
Riemannian and Symplectic geometries. In section $ 3 $, we briefly 
review the construction of Sen entropy function method and thereby 
show its equivalence with the corresponding constrained dynamical system. 
In section $ 4 $, we describe geometric properties of Sen entropy function, 
which correspond to moduli dependent interacting statistical system. 
In section $5 $, we have defined the spectrum and analyzed the convexity 
of Sen entropy function of a given black hole, from the perspective 
of Krull of the underlying algebras. In section $ 6 $,
we apply the above consideration to the case of Calabi-Yau
compactification(s) and provide an algebraic notion to the
deformed S-duality transformations. Finally, in section $ 7 $, 
we present perspective remarks and open issues for a future
investigation.
\section{Real and Complex Geometry}\label{two}
In this section, we provide a brief review of the Riemannian
geometry and its topological connection with the symplectic
geometry. From the perspective of the subsequent sections, we
define a set of needful tools, \textit{e.g.}, existence of closed 
skewsymmetric forms on an even dimensional Riemannian manifold and review 
some further developments, see for example the Refs.\cite{BookGeometry,
BookGeometry1, BookGeometry2}. In particular, we shall offer 
further refinements of the K\"ahler geometry in section $6$. 
Importantly, the analysis of the section $6$ explores details 
of the cases pertaining to Calabi-Yau and deformed S-dualities.
What follows next that we setup the notations and conventions for
a given manifold as follows. Let $ E_n:= \lbrace (y_1,y_2,\ldots,y_n) \vert
y_i \in R \rbrace $ be a set, then $ U \subseteq E_n $ can be
defined as an open ball, if there exists a positive radius $ r $
with a given norm on the set $ U $ such that $ U(y,r):= \lbrace x \in
E_n \vert \Vert x-y \Vert < r, \forall x \in E_n  \rbrace$.
Further, for a given $ n $ dimensional Riemannnian manifold $ \mathcal
M_n $, we shall consider an open ball, which is centered at the
point $ a $ and has a finite radius $ r $, as the set $
B_{\mathcal M_n}(a,r)= \lbrace x \in \mathcal M_n \vert \Vert x-a
\Vert < r \rbrace $. Thus, an open cover of $ \mathcal M \subset
E_s $ can be considered as a finite collection $\lbrace U_\alpha
\rbrace $ of the open sets in $ \mathcal M $, that is endowed in the
union $ \mathcal M:= \bigcup_\alpha U_\alpha $. Subsequently, a
subset $ \mathcal M $ of $ E_s $ is an $ n $-dimensional smooth
manifold, if we are given a finite collection $ \lbrace U_\alpha:
x_\alpha ^1, x_\alpha ^2,\ldots,x_\alpha ^m,\ldots \rbrace $ such
that the following characterizations hold:

$(a)$ The set $\lbrace U_\alpha \rbrace$ forms an open cover of $
\mathcal M $, where $ U_\alpha $ is said to be a co-ordinate
neighborhood of $ \mathcal M $.

$(b)$ We shall assume that each $x_\alpha ^r $ is a $C^\infty$
real valued function with a finite domain $U_\alpha$, i.e. the map
$x_\alpha ^r:U_\alpha \longrightarrow E_n $ is a $C^\infty$-map.

$(c)$ There exists a family of 1-1 maps $x_\alpha :U_\alpha \longrightarrow
E_n $. The 1-1 property of this map shows that the collection 
$x_\alpha(u)= (x_\alpha ^1(u), x_\alpha^2(u),\ldots,x_\alpha ^m(u),\ldots)$ 
offers an existence of the local charts on $\mathcal M$, if the range of the 
mappings are defined over the open sets $W_\alpha \subset E_n $. For a given
family of $C^\infty$-maps $\{x_\alpha^r (u)\}$, $r^{th}$-local co-ordinate
is defined by the local chart $ x_\alpha $. Thus, for an arbitrary
parametrization $ u $ of $\mathcal M$, the coordinate
transformations are regular, if (i) Jacobian determinant
$det(\frac{\partial \overline{x^i}}{\partial x^j})\neq 0 $ and
(ii) mapping class property $x_\alpha:U_\alpha \longrightarrow
W_\alpha \subset E_n$ holds.

$(d)$ Further, if $(U,x^i)$ and $(V,\overline{x^j})$ are two local
charts on the manifold $\mathcal M$ such that $U \bigcap V = \phi
$, then 1-1 nature of the coordinate chart mappings allows us to
express the one set of parameters in terms of the other. In particular, 
we have $x^i=x^i(\overline{x^j})$ with the following inverse set
of the coordinate functions $\overline{x^k}= \overline{x^k}(x^l)$.
Henceforth, we shall assume that the characteristic functions
$x_\alpha^r (u)$ are $C^{\infty}$-maps and thus they form an
admissible set of coordinate functions on the manifold $ \mathcal
M $.

For any compact Riemannian manifold $(\mathcal M,g)$, it follows
that there exists $u:= \bigcup_{\alpha=1}^n U_{\alpha} $ such that
$ \mathcal M \subseteq u $. Moreover, the underlying coordinate
chart can be expressed as the set $\lbrace u_\alpha: x_\alpha^1,
x_\alpha^2,\ldots,x_\alpha^n \rbrace $, where $
x_\alpha^r:U_\alpha \longrightarrow R^1 $ are $ C^{\infty} $-maps $
\forall \alpha \in \lbrace 1,2,\ldots,n \rbrace $. Thus, the
collection $ u $ is a proper open cover, whenever $ \mathcal M = u
$. The transition functions $\lbrace x_\alpha^{-1} \circ x_\beta
\rbrace$ provide us to go back and forth on $\mathcal M $, i.e. we
can go from any chosen $ U_\alpha:= \lbrace x_\alpha^1,
x_\alpha^2,\ldots,x_\alpha^n \rbrace $ to other $
\overline{U}_\alpha:= \lbrace \overline{x}_\alpha^1,
\overline{x}_\alpha^2,\ldots,\overline{x}_\alpha^n \rbrace $, as
long as the transformations preserve $ \Vert \frac{\partial
\overline{x^i}}{\partial x^j} \Vert \neq 0 $. Physically, if
$\mathcal M$ is contained in some finite cover, then we shall say
that $\mathcal M$ is a compact manifold. Further, every Riemannian
manifold $(\mathcal M_n ,g)$ is endowed with the line element $
ds^2:= g_{ij}dx^idx^j $, which defines distance by the standard
inner product structure on the tangent manifold $T_p \mathcal M_n
$ of the manifold $ \mathcal M_n $. Moreover, let $ \mathcal T_1 $
and $ \mathcal T_2 $ be any two topological vector spaces, then
there exists the mapping $ f:\mathcal T_1 \rightarrow \mathcal T_2
$, which turns out to be a homeomorphism, if $(a)$ $f$ is bijective,
$(b)$ both $f$ and $f^{-1}$ are continuous. In particular, let $
\mathcal M_1 $ and $ \mathcal M_2 $ be two such manifolds, then
the map $ f:\mathcal M_1 \rightarrow \mathcal M_2 $ is a
diffeomorphism, if $(a)$ f is homeomorphism, $(b)$ both the
function $f$ and its inverse $ f^{-1} $ are differentiable.
Physically, a square can be viewed as the homeophorphic image of a
circle, which is diffeomorphic to the corresponding ellipse.

As mentioned earlier, we wish to examine the nature of the
embeddings and spectrum of the associated Hessian matrix of 
Sen entropy function of the given black hole. Thus, we recall 
some symplectic geometric properties of an even dimensional Riemannian manifold,
in order to examine the geometric and algebraic properties as per
the interest of the present work. For the given embeddings, the geometry 
of closed skewsymmetric forms is essentially topological in nature, 
and thus we shall often talks about the symplectic topology, 
in due course of the symplectic geometry. For instance, 
one of the common feature of the lowest dimensional symplectic geometry 
is that it is the two dimensional Riemannian geometry, 
which measures the area of the complex curves, instead of 
the length of the real curves. In the case of Euclidean spaces,
the basic notion of the symplectic geometry can be described as per
the following datum. Consider $ R^{2n} $ with the symplectic form $ \omega_0 =
d x_1 \wedge d y_1 + \ldots + d x_n \wedge d y_n $, then there
exists an isomorphism, which could explicitly be given by the
formulae: $ X= \frac{\partial}{\partial x_j}\rightarrow \iota_X
\omega_0= d y_j $ and $ \frac{\partial}{\partial y_j} \rightarrow
-d x_j $. This isomorphism turns out to be a rotation through the
quarter turn, if both the tangent space $ T_X R^{2n} $ and the
corresponding cotangent space $ T_X^{\ast} R^{2n} $ of $ R^{2n} $ 
are identified as: $ \frac{\partial}{\partial x_j} \equiv e_{2j-1}= dx_j $ 
and $\frac{\partial}{\partial y_j} \equiv e_{2j}= dy_j $.

Although, a symplectic form can be regarded as some geometrical
structure in nature, however many problems arise, when one
posits the existence of a symplectic structure, which reduce to
an easier problem in the topology. Thus, the symplectic geometry
puts whole of its emphasis on the topological aspects of the
geometry. Physically, the origin of the symplectic geometry lies
in the fundamentals of classical mechanics and thus the most natural 
example of a symplectic manifold corresponds to the Euclidean phase 
space $R^{2n} $, which is the space parameterizing the positions and
momenta of a classical system with the $ n $ degree of freedoms.
The associated symplectic form is defined by $ \omega_0 = \sum
dp_i \wedge dq_i $, where the position variables of the system are
$ q_i $ and the corresponding momenta are denoted as $ p_i $. In
particular, one of the fundamental attention in the symplectic
geometry is given to the Darboux theorem, which locally connotes
that there exists a set of coordinates, in which the symplectic
two form can be given by the standard Euclidean symplectic form $
\omega_0 $. Moreover, the Moser's method allows, for given any
family of the two forms $ \lbrace {\omega_t; t \in [0, 1]} \rbrace
$ satisfying an appropriate set of hypotheses, \textit{viz.} an existence
of the interpolation between a given symplectic form $ \omega_1 $
and the standard Darboux symplectic form $ \omega_0 $, that one
can construct a family of diffeomorphisms $ \lbrace \varphi_t ; t
\in [0, 1] \rbrace $ such that the following composition holds $
\varphi_t^{\ast} \omega_t= \omega_0 $. Thus, $\forall t \in
[0,1]$, it is possible to exhibit dynamical properties of the
given system with the fact that one can diffeomorphically pull 
back the family $ \omega_t $ to standard Darboux form $ \omega_0 $.

In general, let $ \mathcal M_{2n} $ be a closed even dimensional
compact smooth manifold without boundary. Then, there exists a
smooth symplectic structure $ \omega $ such that $\omega$ is 
closed, i.e. $ d \omega = 0 $ and nondegenerate, i.e., $\omega^n =
\omega \wedge \omega \wedge \ldots \wedge \omega= 0$. For such a 
2-form $\omega$ on $ \mathcal M $, the nondegeneracy condition 
is equivalent to the fact that $ \omega $ induces an isomorphism 
between the vector fields and $1$-forms, as the mappings: 
$ T_X \mathcal M \longrightarrow T_X^{\ast} \mathcal M $, for all 
$ X \longrightarrow \iota_X \omega= \omega(X,.) $. For any $
C^{\infty} $-function $ H: \mathcal M \rightarrow R $ on a
symplectic manifold $ (\mathcal M,\omega ) $, the equation  $ d H(
\cdot )= \omega( X_H , ·) $ defines the associated Hamiltonian
vector field $ X_H $ to the corresponding Hamiltonian $ H $, and
thus the perspective of the quantization. Furthermore, the
associated vector field offers a smooth Hamiltonian function on
the symplectic manifold $ (\mathcal M,\omega ) $. Moreover, there
is a family of contractible Riemannian metrics $\{ g_J \}$ on the
manifold $ \mathcal M $ such that the associated symplectic form $
\omega $ is constructed via $ \omega $-compatible almost
complex structure $ J $, i.e., there exists an automorphism $ J: T
\mathcal M \rightarrow T \mathcal M $ such that $ J^2 = -Id $,
which defines the associated complex vector bundle $ T \mathcal M $. In this
case, the compatibility conditions are: (i) symmetry $ \omega(x, y
)= \omega(J x, J y ) $ and (ii) non-degeneracy $ \omega(x, J x)> 0
, \forall x \neq 0 $. These conditions imply that the associated
bilinear form $ g_J: g_J(x, y )= \omega(x, J y ) $ is the
Riemannian metric on the manifold $ \mathcal M $. Thus, we find,
for all compatible $\omega $, that the set of possible $ J $ is
nonempty and contractible. For a detailed treatment of the above
notions see \cite{ArnoldMcduffSalamon,ArnoldMcduffSalamon1} and
the references therein. Interestingly, one of the main aspect of 
the symplectic structure properties is that there exists its explicit 
connection with the Riemannian geometry and K\"ahler geometry.
Subsequently, in sections $4$ and $5$, we examine these notions 
from the perspective of embedding theory and associated spectra 
of Sen entropy function of the given black hole.

In the due course of our geometric study, we shall illustrate that
there exists the associated notion of generalized K\"ahler manifolds, 
which essentially consist of a pair of commuting generalized complex 
structures. For given complex manifold and the corresponding symplectic 
manifold, the two generic examples of the generalized complex structures 
arise as follow. First, the case of the generalized complex structure 
is simply the standard complex structure, which could be considered as 
the following bundle maps. Let $ \mathcal M $ be an even dimensional 
real manifold, which is equipped with the integrable almost complex
structure $ J: T \mathcal M \rightarrow T \mathcal M $, then the
automorphism $ \mathcal J_J= \left (\begin{array}{rr}
    J & 0 \\
    0 & -J^{\ast} \\
\end{array} \right) $
defines a generalized complex structure on $ \mathcal M $.
Similarly, if $ \omega $ is the standard symplectic structure on $
\mathcal M $, then $\omega$ can be expressed as the following
skew-symmetric map $ \omega: T \mathcal M \rightarrow T^{\ast}
\mathcal M $. Thus, for a given even dimensional manifold $
\mathcal M $, the second class of the generalized complex
structures can be defined as the automorphism group $\mathcal
J_{\omega}= \left (\begin{array}{rr}
    0 & -\omega^{-1} \\
     \omega & 0 \\
\end{array} \right) $.

It is worth mentioning that the above generalized geometric
structures remain well-defined on any even dimensional manifold.
Further generalized geometric consequences pertaining to 
Calabi-Yau and symplectic manifolds are discussed in the 
subsequent sections. From the perspective of the generalized 
Calabi-Yau manifolds, such structures turn out to be either 
of the odd type or even type, which can be transformed by
the actions of underlying diffeomorphisms and thus an existence
of the closed 2-forms on $ \mathcal M $. For the six dimensions
manifolds, it is interesting to note that such generalized
geometric notions can be characterized by the critical points of
the natural variational problem on ceratin closed forms
\cite{HitchinLiYau,HitchinLiYau1}. Subsequently, the moduli space
properties may locally be characterized by a well-defined
consideration of an ensemble of open sets, which are shown
either to be contained in the odd cohomology or in the even cohomology.
In this sense, our analysis anticipates the importance of 
generalized complex structures, which interpolate between the
complex and symplectic structures. Thus, the language of the
present note naturally fits on the interface of the intrinsic
algebraic geometry. In particular, we focus our attention on 
Calabi-Yau embeddings and the associated generalized S-duality
transformations in section $6$. Thereby, the perspectives of
differential geometry and commutative algebra are jointly examined
for Sen entropy function of the extremal black holes.
\section{Sen Entropy Function}\label{three}
In this section, we present a brief review of Sen entropy function
method and thereby offer the attractor entropy of an extremal
black hole. Subsequently, we shall invoke the corresponding
geometric connection from the perspective of constrained
dynamical system. In particular, this section is intended to
provide a set of needful tools for the structures, which emerge as
certain geometric and algebraic properties of Sen entropy function. In
this set up, the extremal black hole shall be viewed as per the
following definition. A four dimensional black hole is said to be 
the extremal solution \cite{SenEntropyFunction,Sen0,FerraraStrominger,
FerraraStrominger1,FerraraStrominger2, FerraraStrominger3}, if (i)
it has $ AdS_2 \times S^2 $ near horizon geometry, which is in particular
known as Robinson-Berttoti vacuum and (ii) the most general
near horizon background fields respect $ SO(2,1) \times SO(3) $
symmetry. In this framework, the entropy of an arbitrary extremal
black hole can be defined as the limit $ S^{ext}_{BH}:=
Lim_{h\rightarrow 0}(S_{BH}) $, where $ h:= r_{+}- r_{-} $ and $
S_{BH} $ are respectively the difference between the outer and
inner radii of the horizon and the entropy of the corresponding
non-extremal black hole. Thus, the extremal limit is reached, when
both the outer radius $ r_{+} $ and inner radius $ r_{-} $ coincide.
Such a limiting procedure is necessary, since an extremal black
hole does not possess bifurcate killing horizon
\cite{Wald,Wald1,Wald2}. Henceforth, we shall assume that $ S_{BH}
$ is well-defined, for a given regular horizon black hole.

Let us consider an arbitrary theory of the gravity coupled with a
set of abelian gauge fields $\lbrace A_{\mu}^{(i)} \rbrace$,
scalar fields $ \lbrace \phi_s \rbrace $ and arbitrary combinations
of their covariant derivatives. Then, under such a consideration. 
Refs. \cite{Wald,Wald1,Wald2} show that the Lagrangian density 
can be expressed as 
$ \mathcal L:= \mathcal L[g_{\mu \nu}, Dg_{\mu \nu},\ldots, \phi_s,
D\phi_s, \ldots, F_{\mu \nu}, DF_{\mu \nu},\ldots, \gamma] $.
Moreover, the Thomos replacement theorem \cite{Wald,Wald1,Wald2}
leads to the fact that the Lagrangian density $ \mathcal L $ can
be written in a manifestly covariant form and thus it remains
independent of the background field $ \gamma $. In particular, let
us focus our attention on those higher derivative theories for
which the covariant derivatives of all tensor fields vanish. Then,
the entropy of the black hole can be computed from Wald formula
\cite{Wald,Wald1,Wald2}. For a given event horizon area $ A_H $,
one finds that Wald entropy of the black hole can be expressed as 
$ S_{BH}= 8 \pi \frac{\partial \mathcal L}{\partial 
R_{\alpha\beta\gamma\delta}} g_{\alpha\gamma}g_{\beta\delta} A_H
$.

In fact, the most general solution of the equations of motion,
which remain consistent with (i) $ SO(2,1) \times SO(3) $ symmetry 
and (ii) $ AdS_2 \times S^2 $ near horizon geometry, takes the 
following form: $ ds^2:= g_{\mu\nu} dx^{\mu} dx^{\nu}= v_1(-r^2dt^2+
\frac{dr^2}{r^2})+ v_2(d\theta^2+ sin^2\theta d\phi^2),\phi_s= u_s
$ and $ \lbrace F_{rt}^{(i)}= e_i, F_{\theta\phi}^{(i)}=
\frac{p_i}{4 \pi} sin \theta \rbrace $, where $ v_1, v_2,\lbrace
u_s \rbrace $ and $\lbrace e_i \rbrace, \lbrace p_i \rbrace $ are
the constants which label the solution of the equations of motion.
In order to do so, let us define the following horizon function $
f(\overrightarrow{u},\overrightarrow{v},
\overrightarrow{e},\overrightarrow{p}):= \int_{S^2} d\theta d\phi
\sqrt{-det g} \mathcal L, \forall (\theta, \phi) \in S^2 $, where
$ \lbrace e_i \rbrace $ are $ (rt) $ components of $
F_{\mu\nu}^{i}:= \partial_\mu A_\nu^{i}- \partial_\nu A_\mu^{i}$.
Further, the electric charges of the theory are measured by $
q_i:= \frac{\partial f} {\partial e_i} $. The $ \lbrace p_i
\rbrace $ denote the magnetic charges of the corresponding black
hole. From the perspective of $ AdS_2 \times S^2 $ near horizon
geometry, we notice further that the function $ f $ is evaluated
as an integral over the horizon, which involves only the angular
coordinates. Thus, Sen entropy function can be defined as Legendre
transform of the above function $ f $ with respect to the electric
variables $\lbrace e_i \rbrace$. In this consideration, Sen
entropy function \cite{SenEntropyFunction,Sen0} can in general be
expressed as $ F(\overrightarrow{u},\overrightarrow{v},
\overrightarrow{e}, \overrightarrow{p}):= 2 \pi
(\overrightarrow{e}. \frac{\partial f(\overrightarrow{u},
\overrightarrow{v},\overrightarrow{e},
\overrightarrow{p})}{\partial \overrightarrow{e}}-
f(\overrightarrow{u},\overrightarrow{v},\overrightarrow{e},
\overrightarrow{p})) $, where $.$ denotes inner product between
near horizon electric fields $\overrightarrow{e}$ and the
corresponding derivative operators $\frac{\partial }{\partial
\overrightarrow{e}}$. Thus, the entropy of the limiting extremal
black hole is obtained as the extremum value of the entropy
function $ F(\overrightarrow{u},\overrightarrow{v},
\overrightarrow{e},\overrightarrow{p})\vert_{extremum
(\overrightarrow{u},\overrightarrow{v})} $. This shows that Sen entropy 
function method \cite{SenEntropyFunction,Sen0} is consistent with 
the equations of the motion. In this sense, it is worth mentioning 
that Sen entropy function method is a more suggestive manner of 
the standard attractor mechanism 
\cite{FerraraStrominger,FerraraStrominger1,FerraraStrominger2,FerraraStrominger3}.

Furthermore, the calculation of the entropy (function) of an
extremal black hole can be generalized to arbitrary D-dimensional
spacetime. To do so, we shall proceed as follows. Let $\lbrace q_i
\rbrace $ be a finite collection of the electric charges
associated with the $D$-dimensional $1$-form gauge fields $\lbrace
A_i \rbrace $ and $\lbrace p_i \rbrace$ be a finite collection of
the corresponding magnetic charges dual to the $( D-3) $-form
gauge fields $\lbrace B_i \rbrace $. Let $ S(\overrightarrow{p},
\overrightarrow{q}) $ be the entropy of the $D$ dimensional black
hole with $ AdS_2 \times S^{(D-2)} $ near horizon geometry. Now,
one can choose a local co-ordinate system such that $ AdS_2 $ part
of the spacetime metric is proportional to $( -r^2 dt^2 +
\frac{dr^2}{r^2})$, then for all diffeomorphism covariant Lagrangian 
density $ \mathcal L[g,\phi_s,...]$, Refs. \cite{Wald,Wald1,Wald2}
show that the entropy of the black holes satisfies $ LT_{\lbrace q_i
\rbrace}(\frac{S_{BH}(\overrightarrow{p}, \overrightarrow{q})}{2
\pi}) = \int_{S^{D-2}} \sqrt{-det g} \mathcal L[g,\phi_s,...] $;
where $LT$ denotes Legendre transformation of the entropy $
S_{BH}(\overrightarrow{p}, \overrightarrow{q}) $. Here, the above
mentioned Legendre transformation is taken with respect to the
electric charges $\lbrace q_i \rbrace$. Physically, for a given
set of electric charges $ \lbrace q_i \rbrace $, the respective
conjugate variables $ \lbrace e_i \rbrace $ represents a set of
radial electric fields, which are associated to $i^{th}$ horizon
valued gauge field of the considered black hole configuration.

In particular, let us consider Robinson-Berttoti $ AdS_2
\times S^{(D-2)} $ near horizon configuration, such that the
parameters $ \lbrace v_1, v_2 \rbrace $ parameterize the
respective sizes of $ AdS_2$ and $S^{(D-2)} $. Then, for the given 
horizon values of electric fields, magnetic fields, scalars fields 
and vector fields of the extremal black hole, the Legendre
transformation of $f = \int_{S^{(D-2)}} \sqrt{-det g} \mathcal L $
with respect to $ \lbrace e_i \rbrace $ is defined such that we
have $ q_i:= \frac{\partial f}{\partial e_i} $. This implies that 
$2\pi \times LT_{ \lbrace e_i \rbrace }(f) = F $. Thus, Sen entropy
function $ F(u_s; v_k; q_i, p_a)$ \cite{SenEntropyFunction,
SenEntropyFunction1,SenEntropyFunction2, SenEntropyFunction3,
SenEntropyFunction4,SenEntropyFunction5, SenEntropyFunction6,
SenEntropyFunction7,SenEntropyFunction8}, as a function of the
charges, near horizon moduli and other parameters, if any, can be
expressed as $ F = 2\pi \times LT_{ \lbrace e_i \rbrace
}(\int_{S^(D-2)} \sqrt{-det g} \mathcal L) $. Consequently, it
follows, from the equations of motion, that the sizes $\lbrace v_k
\rbrace$ of $ AdS_2 $ and $ S^{(D-2)}$ and the near horizon values
$\lbrace u_s \rbrace $ of underlying scalar fields $ \lbrace
\phi_s \rbrace $ can be determined by the extrimization of Sen
entropy function $ F $. For a set of given electric-magnetic
charges of the black hole, the above mentioned extrimization is
performed with respect to both the near horizon moduli $\lbrace u_s
\rbrace $ and parameters $\lbrace v_i \rbrace $. In the other
words, the equations determining the horizon values of $\lbrace
u_s \rbrace $ and $\lbrace v_i \rbrace $ are given by:
$\frac{\partial F}{ \partial u_s} = 0 \Rightarrow u_s = u_s^0 ,
\forall s $ and $\frac{\partial F}{\partial v_i} = 0 \Rightarrow
v_i = v_i^0 , \forall i $. One of the central result of Sen
entropy function method is that the horizon entropy of the
corresponding black hole is given by the attractor fixed point 
formula: $S_{BH}= F(u^0_s;v^0_k;q_i,p_a)$.

From the perspective of the standard attractor mechanism and fixed
point behavior of the scalar fields, Sen entropy function method
provides an efficient tool to incorporate higher derivative
corrections to the attractor equations and thus is an appropriate
framework for the calculation of the entropy of an extremal black
hole. In particular, one can examine the stability properties of
a class of extremal black hole solutions. To do so, let us consider 
$\mathcal N = 2$ supergravity theory with a given Lagrangian density 
$ \mathcal L $, which contains arbitrary finite collection of  
higher derivative covariant tensor fields. Let $f$ be the reduced
Lagrangian density over the horizon of the extremal black hole
such that the Legendre transformation of $f$ with respect to $
\lbrace e_i \rbrace $ provides the associated Wald entropy, as the extremum
of $f$. Further, interesting examples include the followings: (i)
$\mathcal N = 2 $ BPS black holes interacting with Weyl tensor
multiplet and (ii) non-BPS black holes with $ R^2 $-corrections.
In this case, it is known that the higher derivative $
\alpha^{\prime} $-corrections arise precisely due to the
non-holomorphic corrections to the pre-potential
\cite{CarSapna,CarSapna1}.

From the perspective of $ \alpha^{\prime} $-corrections, an
equivalence of Sen entropy function with the corresponding reduced
dynamical system follows from the consideration of the function $
f= \int_{S^{D-2}} d\Omega \mathcal L $, as the reduced Lagrangian
density over the horizon of the black hole. Indeed, the reduced
hamiltonian density can be defined as Sen entropy function $
F= \sum_i e^i q_i- f $, where $ q_i= \frac{\partial f}{\partial
e^i} $ define constraints on the corresponding dynamical system. 
As per the present consideration, $ \lbrace q_i \rbrace $ and 
$ \lbrace e_i \rbrace $ can be respectively treated as co-ordinates 
and momenta, for a given entropy function $ F(p_i, q_i, e_i, v_i, u_s) $. 
In the electric and magnetic pictures, these variables are precisely
related via the respective Legendre transformations $ q_i= \pm \frac{\partial
f}{\partial e^i} $ and the inverse Legendra transformations $ e^i=
\mp \frac{\partial f}{\partial q_i} $, where $ \pm $ signify the
sign conventions being chosen. Moreover, it is not difficult to
see that the magnetic charges $ \lbrace p_i \rbrace $ are
constrained by the Bianchi identities and thus the corresponding
magnetic fields satisfy $ B_i \sim p_i, \forall i $. The variables
$ \lbrace v_i, u_s \rbrace $ define the radii of $ AdS_2 $ and $
S^{D-2} $ and the underlying scalar moduli, whose horizon values
are fixed by Sen entropy function extremization procedure. In
a chosen duality frame of the charges, it thus follows that Sen entropy
function method is equivalent to certain constrained dynamical
system. For example, in the electric description, such a
consideration shows that all the magnetic charges are fixed by
Bianchi identities and thus they remain proportional to the
corresponding magnetic fields.

What follows next is that Sen entropy function facilitates 
us to study the fixed point behavior of scalar fields and other
parameters. As per the definition of the attractor mechanism, 
this enables us to compute the macroscopic entropy of arbitrary
extremal black hole. Subsequently, we shall examine algebraic and
geometric properties of Sen entropy function and show, from the
perspective of constrained dynamical systems, that our framework
exhibits a set of embeddings and convexity relations. In the next
section, we offer such an exposition from the perspective of
embedding theory and finite parameter Sen entropy functions.
\section{Embeddings and Entropy Function}\label{four}
In this section, we study geometric properties of the embeddings
associated with Sen entropy function of an extremal black hole coonfiguration.
Subsequently, we define the local charts, atlas, metric tensor,
generalized symplectic transformations and the associated Legendre
transform on the underlying manifold. From the perspectives of the
local and global geometry, one of the central motivation is to
investigate the natural structures of $ C^{\infty} $ manifold $
\mathcal M $, which possesses a Riemannian like metric leading to
the real and complex geometric structure(s), such that the manifold
$ \mathcal M $ of the real dimensions $ 2n $ can be identified as a
symplectic manifold. Most of the natural structures may be studied by
using the tools of the complex geometry, \textit{viz.} the charts of an
atlas are identified as a finite collection of the open subsets in
$ C^n $. Thus, along a given fiber, we may assume that the
compositions of the charts are described by the holomorphic maps
among the finitely many domains of $ C^n $. Further, the very
natural structures of $ C^{\infty} $ manifold $ \mathcal M $
promote us to define the symplectic structures, which we wish to
explicate for the case of an even dimensional manifold $ \mathcal
M_{2n} $ and thereby show the existence of a nondegenerate close
2-form $ \omega $. For a given $\mathcal M_{2n}$, the classical
geometry over the complex numbers anticipates K\"ahler geometry,
which, as the geometry of a complex manifold, possesses a compatible
Riemannian metric. The concerned algebraic properties have been
relegated to the subsequent section. Notice that the conventional
Riemannian geometry has an extremely rich set of geometric
structures, \textit{viz.} there exist interesting local structures.
However, the corresponding symplectic geometry offers many 
important elements towards the global structures. Both the 
above local and global geometric structures intersect in the 
realm of K\"ahler geometry, where both of the above two geometric 
structures remain compatible. Thus, one can determine the one of 
the structure from the other in a natural way. In particular, 
the compatibility properties constitute the K\"ahler structures 
on the manifold $ \mathcal M $.

For a black hole, we begin our analysis by recalling the fact that 
Sen entropy function $ F(u_s, v_k, q_i, p_j) $ can essentially be
defined as the map $ F:\mathcal M \rightarrow R $. In particular,
for the case of the extremal black holes in $D=4$, we have the 
following manifold $\mathcal M = \mathcal M_\phi \otimes
\mathcal M_2 \otimes \mathcal M_{2n} $. In the case of quantized
charges $ \lbrace q_i, p_j \rbrace $, the base manifold $ \mathcal
M_{2n} $ may be regarded as a symplectic manifold with the
symplectic structure $ \omega^2:= \Sigma_{i<j} \omega_{ij} dp_i
\wedge dq_i $\cite{VGVD,VGVD1,VGVD2,VGVD3, ArnoldMcduffSalamon,
ArnoldMcduffSalamon1}. As per the theory of differentiable manifolds, 
$ \mathcal M_{2n} $ is a symplectic manifold, if there exists 
a differential $2$-form $ \omega $ such that the manifold
$\mathcal M_{2n}$ satisfies the following two properties: (i) the
2-form $ \omega $ is nondegenerate, i.e. the matrix $
\omega_{ij}(\overrightarrow{p}\overrightarrow{q}) $ is invertible;
(ii) the 2-form $ \omega $ is closed, which, with the help of the
exterior differential operator $ d $, may be expressed as $ d
\omega= 0 $. Notice further that the manifold $ \mathcal M $ is
defined as a mixed deformation of $ \mathcal M_{2n} $, 
$ \mathcal M_\phi $ and $ \mathcal M_2 $ in a given order
of the composition, or the converse order. Thus, a consistent 
interpretation of the manifold $ \mathcal M $ may be offered as 
a semisymplectic manifold, whenever the base manifold $ \mathcal M_{2n} $ 
is treated as the symplectic manifold. What follows in the sequel 
that we shall focus our attention on $ D= 4 $ extremal black hole 
entropy functions. This is because the case of arbitrary $ D $ dimensional 
black hole spacetimes arises as an obvious extension $ \mathcal M_{\phi}
\rightarrow \mathcal M_{\tilde \phi } $, where $\tilde \phi$ is a
finite collection of possible fields and other parameters, if any.

In order to offer an insight of the present consideration, let us
first introduce a set of mappings, \textit{viz.} maps from the space
$\mathcal M_{i}$ to the space $\mathcal M_{j}$ such that, for each
pair of indices $\{i,j\}$, the resulting composition $ \mathcal D
$ can be defined as the following embedding $\mathcal M_{2n}
\hookrightarrow^{\mathcal D} \mathcal M$. In order to determine
a well-defined composition, we may notice that there exist the 
two sequences of embeddings $\lbrace \mathcal D_1 \rbrace$ and
$\lbrace \mathcal D_2 \rbrace$, which are respectively associated
to the horizon values of the moduli $ \lbrace \phi_i \rbrace $ and
parameters $ \lbrace v_i \rbrace $, for a given set of charges. 
For this proposition, we may define an embedding $ \mathcal D $ as
the extension map $(\overrightarrow{p},\overrightarrow{q}) \mapsto
(\overrightarrow{p},\overrightarrow{q},
\overrightarrow{0},\overrightarrow{0})$. Thus, there exists a
translation map $ T: \mathcal M \rightarrow \mathcal M $ such that
$ (\overrightarrow{p},\overrightarrow{q},
\overrightarrow{0},\overrightarrow{0}) \mapsto
(\overrightarrow{p},\overrightarrow{q},
\overrightarrow{u},\overrightarrow{v})$, for some $ \phi_{i} \in
\mathcal M_{\phi}, v_i \in M_{2} $. Moreover, it is not difficult
to show that $ T $ is a diffeomorphism on $\mathcal M$. To examine
the above mentioned geometric properties, let us consider a finite
set of the embeddings $\{\mathcal D_1, \mathcal D_2\}$, and
thereby an explicit definition emerges as the following embedding
sequence $\mathcal M_{2n} \hookrightarrow_{\mathcal D_1} \mathcal
M_{2n} \otimes \mathcal M_{\lbrace \phi_i \rbrace}
\hookrightarrow_{\mathcal D_2} \mathcal M_{2n} \otimes \mathcal
M_{\lbrace \phi_i \rbrace}\otimes \mathcal M_{\lbrace v_i \rbrace}
$.  For a given pair $\{ \mathcal D_1, \mathcal D_2 \}$, there exists
a local isomorphism $ \sim $, which as the complete embedding of $
\mathcal D $, furnishes the following well-defined composition $
\mathcal D:= \mathcal D_2 \circ \mathcal D_1 $. As shown in
table $1$, we notice, from the perspective of a given 
sequence of embeddings on the manifolds $\mathcal M_{\lbrace
\phi_i \rbrace}$ and $\mathcal M_{\lbrace v_i \rbrace}$, that the
well-definiteness property corresponds to the requirement that the
embeddings $ \lbrace \mathcal D_1,\mathcal D_2 \rbrace $ are
defined as per the following compositions: (i) $ \mathcal D_1:=
\mathcal D_{1m} \circ \cdots \circ \mathcal D_{12} \circ \mathcal
D_{11} $ and (ii) $ \mathcal D_2:= \mathcal D_{22} \circ \mathcal
D_{21} $. For a given composition map $ \circ $, it is easy to
show that the other composition $ \mathcal D_1 \circ \mathcal D_2
$ is ill-defined and is not the same as the composition $\mathcal
D= \mathcal D_2 \circ \mathcal D_1 $. This amounts to the fact
that the inverse orientation embedding $ \mathcal D^{\dagger} \neq
\mathcal D $ prefers an order of the compositions and thus the
resulting composition $ \mathcal D $ turns out to be a sequence of
finitely many unidirectional embeddings. It is worth mentioning
that such an underlying orientation does not make a diffident
physical importance, whenever there exists a local coordinate 
frame on the manifold $ \mathcal M $.

\begin{table}
\scriptsize
\centering
\begin{tabular}{c  c}
\\
$\mathcal M_{2n}  \longrightarrow $        &
$\mathcal M_{2n} \otimes \mathcal M_{\lbrace \varphi \rbrace}
\otimes \mathcal M_{\lbrace v \rbrace}$ \\
$ \downarrow_{\mathcal D_{11}}$   &   \\
$\mathcal M_{2n} \otimes R^1_{\lbrace \varphi_1 \rbrace}$ &  $ \wr $    \\
$\downarrow_{\mathcal D_{12}} $   & $\mathcal M_{2n} \otimes
\mathcal M_{\lbrace \varphi_i \rbrace} \otimes R^1_{v_1} \otimes R^1_{\lbrace v_2 \rbrace }$ \\
$\mathcal M_{2n} \otimes R^1_{\lbrace \varphi_1 \rbrace}\otimes R^1_{\lbrace \varphi_2 \rbrace} $ &   $ \uparrow_{\mathcal D_{22}} $ \\
$\vdots $ & $\mathcal M_{2n} \otimes \mathcal M_{\lbrace \varphi_i \rbrace}\otimes R^1_{\lbrace v_1 \rbrace}$ \\
$\downarrow_{\mathcal D_{1m}} $  &   $   \uparrow_{\mathcal D_{21}} $         \\
$ \mathcal M_{2n} \otimes R^1_{\lbrace \varphi_1 \rbrace}\otimes \cdots \otimes R^1_{\lbrace \varphi_m \rbrace}  \sim $
&  $ \mathcal M_{2n} \otimes \mathcal M_{\lbrace \varphi_i \rbrace} $\\
\\
\hline
\end{tabular}
\caption{Sequence of embeddings associated with $\mathcal
M_{\lbrace \varphi_i \rbrace}$ and $\mathcal M_{\lbrace v_i
\rbrace}$.} \label{tab2}
\end{table}

We shall now define an associated Euclidean structure on the
manifold $ \mathcal M $. Let us consider the symplectic system $
\lbrace \overrightarrow{p},\overrightarrow{q} \rbrace $, moduli
fields $ \overrightarrow{\phi} $ and $ AdS_2 \times S^2 $
parameters $ \overrightarrow{v} $. As pointed out in Ref.
\cite{Sen0}, we shall take $ \lbrace \overrightarrow{e_{p_i}},
\overrightarrow{e_{q_i}}, \overrightarrow{e_{\phi_i}},
\overrightarrow{e_{v_i}}\rbrace $ on the same footing. Therefore,
we can define a vector $ \overrightarrow{s}:= \sum_i (p_i
\overrightarrow{e_{p_i}}+ q_i \overrightarrow{e_{q_i}} + u_i
\overrightarrow{e_{\phi_i}} + v_i \overrightarrow{e_{v_i}})$ such
that the semi-Euclidean structure is given by the product
$(\overrightarrow{s},\overrightarrow{s}): = \sum_{i,j} (p_i J_{ij}
q_j + u_i u_{ij} u_j + v_i v_{ij} v_j)$. At this juncture, it is
worth mentioning that the above geometric notions lie in the
structures of the $J_{ij}$. Thus, it is natural to associate 
an interpretation of the semi-Euclidean structure on the
manifold $ \mathcal M $, with the fact that the symplectic
structures pertain to the elements of the matrix $ J_{ij} $. In
general, the associated $ J_{ij} $ may or may not imply an
uniquely defined Euclidean structure. Considering the manifold
$\mathcal M$ as the deformation of an even dimensional Riemannian
manifold $\mathcal M_{2n}$, we can determine the semi-Euclidean
structure by defining the following line element $ ds^2:=
\sum_{i,j} b_{ij}ds^i ds^j $. After a set of local coordinate
transformations on $\mathcal M$, we can deduce that the metric
tensor $ b_{ij}$ takes the following form $ b_{ij} =\left
(\begin{array}{rr}
    0 & -E \\
     E & 0 \\
\end{array} \right)
\bigoplus u_{ab} \bigoplus v_{\alpha\beta}$. In the above
consideration, we find that the  deformed metric tensor appears as
the direct sum of the associated metric tensors pertaining to the
electric-magnetic charges $ \lbrace p_i, q_i \rbrace $, near
horizon values of the scalar moduli $ \lbrace u_i \rbrace $ and
$\{AdS_2,S^2\}$ radial parameters $ \lbrace v_1, v_2 \rbrace $.
Moreover, we see, for finitely many electrically magnetically
charged black holes \cite{Moore1}, that the standard symplectic
matrix $ J= \left (\begin{array}{rr}
    0 & -E \\  E & 0 \\  \end{array} \right) $
reduces to the Gram matrix $ Q_{ab} = \left (\begin{array}{rr}
    p^2 & -p.q \\  -p.q & q^2 \\  \end{array} \right) $.
Thus, for a proper symplectic basis, we observe that the above
mentioned manifold $ \mathcal M $ is locally spanned by a finite 
sequence $ \lbrace s_i \rbrace_{i=1}^{2n+m+2} $, whose components 
are defined as the union of $ 2n $ electric-magnetic charges, 
$ m $ scalar fields $ \lbrace \phi_i \rbrace_{i=1}^{m} $
with near horizon values $ \lbrace u_i \rbrace_{i=1}^{m} $ and the
two near horizon $ AdS_2, S^2 $ spacetime parameters $ \lbrace v_i
\rbrace $. As an artifact of the two dimensional manifold $
\mathcal M_2 $, it is easy to see that the metric on $ \mathcal
M_2 $ can be reduced to the following diagonal metric: $
v_{\alpha\beta} \sim \eta_{\alpha\beta}e^{\Phi} $. In this case,
the resulting function $ \Phi(v_1,v_2) $ may be considered as the
dilaton field on the manifold $ \mathcal M_2 $. In general, for a
given entropy function, it follows that there exists a class of
semi-euclidian structures on $ \mathcal M $, which are locally
spanned by the basis set $ \lbrace
\overrightarrow{e}_{\overrightarrow{p}},
\overrightarrow{e}_{\overrightarrow{q}},
\overrightarrow{e}_{\overrightarrow{\phi}},
\overrightarrow{e}_{\overrightarrow{v}}\rbrace $. Herewith, the
above type of semi-structure may be defined as the product $ (
\overrightarrow{s_1}, \overrightarrow{s_2} ) = (J
\overrightarrow{\xi}, \overrightarrow{\eta}) +
(\overrightarrow{\psi_1},\overrightarrow{\psi_2}) + (w_1,w_2) $,
where $ \overrightarrow{\xi}, \overrightarrow{\eta} \in \mathcal
M_{2n}; \overrightarrow{\psi_1},\overrightarrow{\psi_2} \in
\mathcal M_{\phi} ; w_1,w_2 \in \mathcal M_2 $. Notice further
that the automorphism $ J:R^{2n} \rightarrow R^{2n} $ is the
standard symplectic operator, which may thus be explicitly given
by the matrix $ \left (\begin{array}{rr} 0 & -E \\  E & 0 \\
\end{array} \right) $, where $ E $ represents $ n \times n $ identity
matrix, as in the usual Euclidean cases. Furthermore, it is follows that 
we have $ J^2 = - E_{2n} $. Herewith, from the definition of the
Euclidean structure on $\mathcal M_{2n}$, it is worth mentioning
that the associated Poisson structure on $\mathcal M_{2n}$ can be
expressed as the following bracket $ [ \overrightarrow{\xi},
\overrightarrow{\eta}]_{PB} = (J \overrightarrow{\xi},
\overrightarrow{\eta}) $. From the fact that the left hand side of
the above bracket is symplectic, whereas the right hand side is
symmetric and thus it follows that the operator $ J $ must be
symplectic. Observe further that the standard complex structure on
$\mathcal M_{2n}$ is defined by the map $ z_k \mapsto p_k + i q_k
$ and thus we can define a deformed complex structure by an
extension map $ z_k \mapsto_{\iota} p_k + i q_k + u_k + v_k $,
which could either be identified as the extension $ p_k \mapsto
p_k + u_k + v_k $, or $ q_k \mapsto q_k + u_k + v_k $ or an
arbitrary (linear) combination of the charges $ \lbrace p_k, q_k
\rbrace $ such that the map $ \iota $ remains well-defined. In
particular, the map $ \iota $ defines an embedding: $ \mathcal M_n
\hookrightarrow \mathcal M_n \otimes \mathcal M_m \otimes \mathcal
M_2 $, which, for example, in the magnetic frame could be given by
the extension map $ p_k \mapsto^{\iota} p_k + u_k + v_k $. Thus,
it is easy to see that the corresponding projection map $ \rho $
must be given by the restriction $ \mathcal M_n \otimes \mathcal
M_m \otimes \mathcal M_2\longrightarrow^{\rho} \mathcal M_n $. In
turn, the fixed horizon behavior of the black hole may be defined 
by the composition of translation and restriction maps as the following 
sequence $ (p_k,u_k,v_k) \rightarrow_{T^{-1}} (p_k,0,0)\rightarrow_{\rho}
(p_k)$. Under the above proposition of generalized complex
structure, we may note that there always exists a possibility of
some generalized electric-magnetic duality transformations. For a
given pair of maps $\{T, \rho\}$, the above consideration leads
to the fact that the near horizon configuration of extremal black
holes picks up a duality frame, which could be either in the
electric sector or in the magnetic sector or in a mixed sector
defined as the (linear) combinations of the electric-magnetic
charges $ \lbrace p_k, q_k \rbrace $.

Herewith, we extend our analysis towards the perspective of 
dynamics and stability theory. In order to do so, let us restrict
our attention the general linear mappings, namely, an algebraic group $
GL(n, K)$, where $K$ is a finite field, ring or any algebraic
object, satisfying the symmetry properties of the metric $b_{ij}$.
To concentrate on the geometric physics of an extremal black hole,
we shall offer the analysis when $K$ is a finite field. In the
case when $K=C$, such a consideration may be defined by the following
sequence of mappings $ \lbrace f_k \in C \rbrace $ such that $
GL(n,C):= \lbrace f \vert s_k \rightarrow^{f_k} s_k; k=1,...,n;
\rbrace $. In turn, this perspective allows us to define the
unitary transformation preserving Hermitian scalar product: $
\prec \overrightarrow{s}_1,\overrightarrow{s}_2 \succ:=
(\overrightarrow{s}_1, \overrightarrow{s}_2) +i
[\overrightarrow{s}_1,\overrightarrow{s}_2] $; where the scalar
and skew scalar products are respectively given by the real and
imaginary parts of the above scalar product on $ \mathcal M $. Note that
the complex structure, which as defined earlier with the injection
$ \iota $, is the same as the one defined with the above scalar
and skew scalar products. In order to consider the stability
conditions on $ \mathcal M $, let us first consider the notion that
the usual mechanical stability transformations are defined as an
automorphism, namely, the dynamics on $\mathcal M$ is said to be
stable, if $ \forall \epsilon >0$, there exist (i) $\delta >0 $
and (ii) a map $ g: \mathcal M \rightarrow \mathcal M $ such that
$ \Vert \overrightarrow{s} \Vert < \delta \Rightarrow \Vert g^N
\overrightarrow{s} \Vert < \epsilon, \forall N > 0 $. Under the
generalized or deformed symplectic transformations, it is worth
mentioning that the stability may be defined as an extension of
the above stability transformation on $ \mathcal M_{2n}$. In fact, 
the required extension can be defined by a deformation of the above
symplectic transformation $ g $ and thus a generalized
transformation is said to be ``strongly stable'', iff every
generalized symplectic transformation $ \overline{g} $, which is 
sufficiently close to $ g $, is stable. This shows that there exists a map $
\overline{g} $ such that it remains sufficiently close to $ g $,
if the matrix elements of the map $ \overline{g} $ differs from
that of the map $ g $ by less than a sufficiently small number $
\epsilon $, in a fixed local basis $ \lbrace s_i
\rbrace_{i=1}^{2n+m+2} $ on $\mathcal M$. Thus, the stability of
an arbitrary $ \overline{g} $ on $ \mathcal M $ can be examined in
a preferred local basis $ \lbrace s_i \rbrace_{i=1}^{2n+m+2} $.
Equivalently, the above analysis amounts to the fact that the
restriction map $\rho$ and translation map $T$ on the
corresponding matrix elements $b_{ij}$ reveal the geometric nature
of the black hole horizon configuration. In particular, we find
that the stability of Sen entropy function is given by
$\vert \overline{g}_{ij} \vert \simeq \vert g_{ij} \vert \pm
\epsilon $. Thus, the strong stability of a semisymplectic
transformation relates its coefficients to the given local
transformations on a stable manifold $\mathcal M$.

Let us now construct the generalized symplectic atlas of a deformed
symplectic manifold $ \mathcal M $. Such a construction can
essentially be realized by considering the metric $ b_{ij} $. This
follows from the fact that every symplectic manifold has some
local coordinate basis $ \lbrace \overrightarrow{p},
\overrightarrow{q} \rbrace $. Now, the Darboux theorem offers 
standard symplectic $2$-form $ \omega^2 := d \overrightarrow{p}
\wedge d \overrightarrow{q} $, from which various symplectic
structures may be obtained accordingly. As per say, the definition
of every manifold induces a compatibility condition on the
underlying local charts of the atlas of the manifold. This gives a
condition on the connecting maps $ \lbrace s_i^{-1} \circ s_j
\rbrace $, in order to go from one chart to the other, in a
well-defined way. In order to make coordinate transformations from
a given region to the another, the intersection maps need to
be well-defined, in certain patches of the coordinate space.
In the other words, we can define an atlas of the manifold $
\mathcal M $ with the norm: $ b^2 = \omega^2 + \mu^2 + \nu^2 $,
which may locally be accompanied by the introduction of the
corresponding Euclidian coordinate space $R^{2n+m+2}= \lbrace
((\overrightarrow{p} \overrightarrow{q}); \overrightarrow{u};
\overrightarrow{v}) \rbrace $ and thereby the norm $ b^2$ 
defines nature of the transformation(s) from the one point 
to the other. Such a generalized symplectic structure may be 
realized as the generalized canonical transformations, which are defined 
as $ b^2$ preserving maps. In particular, one needs to solely fix the
compatibility structures of the composition maps $U_{ij}:= \lbrace
s_i^{-1} \circ s_j \rbrace $, for some given $ \omega^2 := d
\overrightarrow{p}\wedge d \overrightarrow{q} $, $ \mu^2=
d\overrightarrow{\mu} \otimes d\overrightarrow{\mu} $ and $ \nu^2=
d\overrightarrow{\nu} \otimes d\overrightarrow{\nu} $. Here, the
above products respectively correspond to the following
matrices $ \left (\begin{array}{rr} 0 & -E \\  E & 0 \\
\end{array} \right) , (u_{ab})_{n \times n} $ and $(v_{\alpha
\beta})_{2 \times 2}$. As defined before, the case of $ \mathcal
M(R) $ follows from the following expansions $b^2= dp_1 \wedge
dq_1+ dp_2 \wedge dq_2 + \ldots + dp_n \wedge dq_n + du_1 \otimes
du_1 + du_1 \otimes du_2 + \ldots + du_1 \otimes du_m + \ldots +
du_m \otimes du_m + dv_1 \otimes dv_1+ \ldots + dv_2 \otimes
dv_2$. In the case when the effective actions are allowed to be
complex, it is worth mentioning that the stability analysis needs
an extension on the corresponding complex manifold $ \mathcal M(C) $.
From the viewpoint of the complex analysis, the well-definiteness
of the resulting complex embeddings requires that the composition
maps $ \lbrace s_i^{-1} \circ s_j \rbrace $ should factorize into
the analytic and antianalytic sectors.

In order to formulate algebraic properties of the above
mentioned geometry, let us focus on the fact that the symplectic
structure defines an effective hamiltonian dynamics. As mentioned
in section $2$, an immediate goal could be to give the precise
local and global stability interpretations on $\mathcal M$, for a
given Sen entropy function $ F(s) $. Subsequently, we shall return to
the associated algebraic definition of Legendre transformation
in the next section. Before describing the details of the
algebraic geometric analysis, it will be prudent to note that $ F
$ is said to be strictly convex, if the Hessian matrix $(d^2 F)_p
(s) \gg 0 , \forall p \in \mathcal M $. Let $ F $ be such a
strictly convex function on $ \mathcal M $. Then, for a given $
l \in \mathcal M^{\star} $, there exists a map $ F_l: \mathcal M
\rightarrow R $ such that the cotangent structure can be defined
by $F_l(w):= F(w)- l(w)$. It is worth mentioning that the Hessian
matrix of the function $ F $ is a quadratic form on $ \mathcal
M $ and thus it can be locally expressed as $(d^2 F)_p (s):=
\sum_{i,j} \frac{\partial^2 F(p)}{\partial w_i
\partial w_j} s_i s_j$. Such a consideration offers the local
definition of the Hessian, which is defined in terms of the basis
functions of the manifold $ \mathcal M $. Notice further that the
corresponding global definition is given by $(d^2 F)_p (s):=
\frac{d^2}{dt^2} F(p+ts)\vert_{t=0}$. Now, it can be easily
verified that both the above frameworks have the same realization
of the dynamical stability. In particular, it follows from
the fact that $(d^2 F)_p = (d^2 F_l)_p \Rightarrow F $ is strictly
convex $ \Leftrightarrow F_l $ is strictly convex. It is worth
mentioning that the determinant of the Hessian matrix of the
considered entropy function $ F(s) $ remains unchanged under the
transformations $b_{ij} \rightarrow b_{ij}- c$, where $c$ is an
arbitrary constant. Herewith, we find that the stability of
arbitrary entropy function $ F(s) $ can be defined as the set $
S_F:= \lbrace l \in \mathcal M^{\star} \vert \ F_t $ is
stable$\rbrace$. In this case, we see that the stable set $ S_F $
is an open and convex subset of the underlying cotangent manifold
$ \mathcal M^{\star} $.
For such $ \mathcal M $, a suitable definition of the Legendre
transform can be offered as follows. For a given entropy function
$ F \in C^{\infty}( \mathcal M, K ) $, the generalized Legendre
transformation is the map $ L_F: \mathcal M \rightarrow  \mathcal
M^{\star} $ such that $ p \rightarrow dF_p \in T_p ^{\star}
\mathcal M \simeq \mathcal M^{\star}; \forall p \in \mathcal M $.
In this case, it is not difficult to verify that the map $ L_F $
is $ (a) $ 1-1, $ (b) $ onto and $(c)$ both $ L_F$ and $L_F^{-1}
$ are continuous. Thus, we find that the map $ L_F $ locally
results into a homeomorphism. From this perspective, we see that
the homeomorphism $ L_F $ becomes a diffeomorphism onto $ S_F $,
whenever $ L_F $ is a differentiable map. In particular, it can be
easily shown that the following conclusion holds. Let $ F
$ be a strictly convex function, then the map $ L_F: \mathcal M
\rightarrow S_F $ is an isomorphism. In fact, $ L_F $ turns out to
be a diffeomorphism onto $ S_F $. Furthermore, the inverse map $
L_F^{-1}: S_F \rightarrow \mathcal M $ describes the unique
minimum point $ p_l \in \mathcal M $ of $ F_l = F - l, \forall l
\in S_F $. Thus, it is easy to see that the point $ p $
corresponds to the unique minimum of $ F(w) - dF_p (w) $. We wish
to remark that the dual space $ \mathcal M^{\star} $,
corresponding to the given $ F $, is completely described by the
map $ F^{\star}: S_F \rightarrow R $. For some $l \in \mathcal
M^{\star}$, the minimum value of $F(s)$ may be defined as $
F^{\star}(l)= -min_{p \in \mathcal M} F_l(p)$. Finally, it turns
out that the duality relation between $ L_F^{-1}$ and $
L_{F^{\star}} $ are expressible in terms of the orthonormal basis
functions of $ \mathcal M$ and $ \mathcal M^{\star}$. In
particular, let $ \lbrace s_I \rbrace $ be the tangent basis and $
\lbrace \tilde{s}^I \rbrace $ be the corresponding cotangent basis
of a given $ \mathcal M $. Thus, as per the general theory of
local vector spaces, we find that the underlying basis functions
satisfy the following orthonormality relation $ s_I \cdot
\tilde{s}^J = \delta_I^J $. Furthermore, the above consideration
implies that we have $ s_I^{-1}= \tilde{s}_I, \forall I $, and
hence the isomorphism property follows as the identification $ L_F^{-1} =
L_{F^{\star}} $. Subsequently, it follows that $ L_F $ and $
L_{F^{\star}} $ are dual to the each other. In the next section,
we shall analyze the above properties of Sen entropy function from
the perspective of commutative algebra and thereby setup the
notions for the Calabi-Yau geometry.
\section{Spectrum and Convexity}\label{five}
As mentioned in section $4$, we can determine a finite set of
embeddings associated with the geometric perspective of $\mathcal
M$. The corresponding Hessian matrix of the entropy function can
be expressed in terms of the metric $ \omega_{ab} $, for a given
set of the electric-magnetic charges, $ u_{ab} $ corresponding to
the scalar fields obtained for a chosen compactification and the
variables $ v_{ab} $ as the gravity parameters characterizing the
spacetime metric tensor of the consideredd black hole. The geometry 
thus obtained from the Hessian matrix of the entropy function fits in
the general framework of the generalized geometry. As mentioned in
section $2$, such a geometry may be reduced to the symplectic
geometry, with a set of associated complex structures $ J:=
\lbrace J^{(i)} \rbrace $ for the case of an even number of the
scalar fields. Thus, there exists an even dimensional Riemannian
manifold whose metric structure can be defined as the bilinear map: 
$ g_J:g_J(p,q) \rightarrow \omega(p,Jq) $. The case
of the odd dimensional $ \mathcal M $ corresponds to an odd number
of the scalar fields. For every given Hessian matrix $ b_{ij} $,
it follows in both the above cases that there exists a Riemannian
manifold $ \mathcal M $. Further, the associated geometry
corresponds to a higher dimensional symplectic manifold $ \mathcal
M $, iff the dimension of the considered manifold $ \mathcal M $
is an even integer.

In this section, the goal is to describe the general procedure for 
determining the algebraic properties, which arise from the Hessian 
quadratic form and associated polynomial function of a given Sen 
entropy function. We begin with the postulates and notions of the
basic commutative algebra, see Refs. \cite{algebra,algebra1,algebra2}
for more details, in order to properly deal with the concepts of the 
spectrum and convexity of the underlying geometry of a given entropy 
function $F(\overrightarrow{s})$. From this perspective, let Sen entropy 
function be defined as the map $ F:\mathcal M \rightarrow R $ such that 
$ F(\overrightarrow{s}) = a $. Then, the level set, as the inverse of a 
singlton $ b $, may be defined as a finite set $ F^{-1}(b)= \lbrace \overrightarrow{s}
\in \mathcal M \vert F(\overrightarrow{s}) = a \rbrace $. For the
reason which follows subsequently, we shall work with an
assignment that there are finitely many co-ordinates $ s_i \in
\mathcal M$. For a given index $ \Lambda $, let $ s= \lbrace s_i
\rbrace $ be a bounded sequence, then the infinite norm of the
vector $ \overrightarrow{s} $ may be defined as the following set
$ \Vert s \Vert_{\infty}= Sup \lbrace \vert s_i \vert : i \in
\Lambda \rbrace $. Moreover, consider $ C( \mathcal M_1,\mathcal
M_2 ) \ni T: \mathcal M_1 \rightarrow \mathcal M_2 $, then the
suppremum norm of such an operator $ T $, for the given norms
$\Vert . \Vert_1 $ on $ \mathcal M_1 $ and $\Vert . \Vert_2 $ on $
\mathcal M_2 $, may be given by $ \Vert T \Vert_{\infty}= Sup
\lbrace \Vert Tu\Vert_2 \vert \Vert u \Vert_1 \leq C \rbrace $. On
the other hand, a submanifold $ \mathcal N \subseteq \mathcal M $
turns out to be compact, if for a given open cover $ u =
\bigcup_{\alpha \in \Lambda} U_{\alpha} $ of $ \mathcal M$, there
exists some $\Lambda$ such that $\lbrace \alpha_i \rbrace_{i=1}^{n} 
\subseteq \Lambda $. In this case, it follows hereby that the submanifold 
$ \mathcal N $ is finitely covered, \textit{viz.} $ \mathcal N 
\subseteq \bigcup_{\alpha = 1}^{n}U_{\alpha} $.

We shall now turn our attention on the analysis of the spectrum
and convexity of Sen entropy function $ F:\mathcal M \rightarrow R $
of the associated black hole. This may most conveniently be
accomplished by noticing the fact that there exists the quadratic
function $(d^2 F)_p (s):= \sum_{i,j=1}^r \frac{\partial^2
F(p)}{\partial w_i \partial w_j} s_i s_j \equiv \sum_{i,j=1}^r
b_{ij}s_i s_j $; where $ r= dim( \mathcal M ) $. Let us consider
the matrix $ \mathcal B:= (b_{ij})_{r \times r} $, then the
stability of the corresponding black hole system may be described by 
the spectrum of the eigenvalues $ \lbrace \lambda_1, \lambda_2, \ldots
,\lambda_r \rbrace $ of $ \mathcal B$. In this interpretation, the
convexity of $ F $ may be realized in the minimal subfield $ K^n
$, containing the sequence $ \lbrace \lambda_i \rbrace_{i=1}^r $
with a finite completion $ \lbrace \tilde{\lambda}_i
\rbrace_{i=1}^{\tilde{r}} $. In such cases, one finds that an
example of the field $ K^n $, associated with black hole entropy
function, turns out to be the real field $ R^n $ for the
configurations whenever the corresponding matrix $ \mathcal B $
ceases to a real symmetric matrix. The case of $ C^n $ may be
similarly extended with the fact that $ \mathcal B $ belongs 
to a set of complex hermitian matrices.

From the perspective of the function theory, let us consider a
field $ K \subset K^n $ as the bounded level set of the (extremal)
black hole entropy function $ F(u_\alpha, v_\beta, p_i, q_j) $.
Then, under the above consideration, Ref. \cite{topalg} shows that 
there exists a convex polynomial hull $ \wp(K) $, which may in turn by 
defined by the set $ \wp(K)= \lbrace \overrightarrow{\lambda}=(\lambda_1,
...., \lambda_n) \in K^n \vert  \vert P(\overrightarrow{\lambda})
\vert \leq \Vert P \Vert_K \forall $ polynomials $
P(\overrightarrow{\lambda}) $ over $ K^n \rbrace $. Here, we have
taken the norm of the polynomial $P$ as the set $ \Vert
P \Vert_K= Sup\lbrace \vert P(\overrightarrow{\mu})\vert \vert
\overrightarrow{\mu} \in K \rbrace $. In general, this allows that
such a polynomial convex hull of $ K $ satisfies $ \wp(K)
\supseteq K $. In particular, for the case of $ \wp(K) = K $, it
may easily be shown that the set $ K $ is polynomially convex.
Herewith, it is easy to see that $ \wp(K) $ has following
properties: (i) $ \wp(K) $ is compact. Considering the fact that $
\wp(K) $ is closed, let $ \varepsilon_j $ be a finite set of
polynomials satisfying the following relation $
\varepsilon_j(\lambda_1, \lambda_2,..., \lambda_n) = \lambda_j $.
By taking $ P = \varepsilon_j $ into the definition of $ \wp(K) $,
it follows, for the case of $ \overrightarrow{\lambda} \in \wp(K)
$, that we have $ \vert \lambda_j \vert \leq Sup \lbrace \vert
\mu_j \vert : \overrightarrow{\mu} \in K \rbrace = C_j < \infty $.
In turn, this implies that $ \wp(K) $ is bounded and thus compact.
In particular, this implies that any p-convex $ K $ defined as the
set $\lbrace \lambda_i \rbrace$ is thus compact. (ii) From the
definition of $p$-convexity, we may further see that the set $K$
remains $p$-convex, iff there exists a polynomial $ P_0 $ such
that $ \vert P_0(\overrightarrow{\lambda^0}) \vert
>\Vert P_0 \Vert_K, \forall \overrightarrow{\lambda^0} \in K^n
\setminus K $. (iii) Moreover, if $K$ is $p$-convex and $ C > 0 $,
then one finds that there exists a polynomial $ Q $ such that $
\Vert Q \Vert_K \leq C, \forall \overrightarrow{\lambda^0} \in K^n
\setminus K $, where $ C <\vert Q(\overrightarrow{\lambda^0})\vert
$. From the consideration of (ii), it may easily be seen with an
appropriate choice of $ P_0 $ that there exists a $ \overrightarrow{\lambda^0}
\in K^n \setminus K $ such that $ K $ is $p$-convex. Such a
consideration leads to the statement that the polynomial $ Q $ 
can be identified as $ Q(\overrightarrow{\lambda})= C P_0(
\overrightarrow{\lambda^0})/\Vert P_0 \Vert_K $.

On the other hand, from the theory of Banach algebra of several
complex variables, Refs. \cite{topalg, BanachAlgebra} make known 
that a compact set $ K \in C $ is $p$- convex, iff $ C \setminus K $ 
is connected. Thus, in order to consider the convexity of a function
of several complex variables, let us define polydisc $ P_{
\overrightarrow{\xi}} $ of polyradius $ \overrightarrow{\xi} =
(\xi_1, ..., \xi_n) $ as the set $ P_{\overrightarrow{\xi}} =
\lbrace \overrightarrow{\lambda} \in K^n \mid \vert \lambda_j
\vert \leq \xi_j,\forall j=1,...,n \rbrace $. Then, the polysubset
$ \Pi $ of $ P_{\overrightarrow{\xi}} $ is a p- polyhedron in $
P_{\overrightarrow{\xi}} $, if there exist polynomials $ \lbrace
P_i \rbrace_{i=1}^m $ such that $ \Pi = \lbrace
\overrightarrow{\lambda} \in P_{\overrightarrow{\xi}} \mid \vert
P_i(\overrightarrow{\lambda}) \vert \leq 1 \forall i=1, ..., m
\rbrace $. Now, one may observe that the followings hold: (i)
Suppose $ m= n$ and $P_j= \xi_j^{-1} \varepsilon_j $, where $
\varepsilon_j $ are the polynomials defined by $
\varepsilon_j(\overrightarrow{\lambda})= \lambda_j $, then $
P_{\overrightarrow{\xi}} $ is a $p$-polyhedron. (ii) There exists
a relation between $p$- convexity and $p$- polyhedron, namely, suppose $
\overrightarrow{\lambda^0} \notin K^n \setminus \Pi $ then either
we have some $\vert \overrightarrow{\lambda^0_j}\vert
> \xi_j $ or there exist some polynomials satisfying
$ \vert P_k(\overrightarrow{\lambda^0})\vert >1 $. In the first
case, we find that $ \vert
\varepsilon_j(\overrightarrow{\lambda^0})\vert = \vert
\overrightarrow{\lambda^0_j} \vert > \xi_j \geq \Vert
\varepsilon_j \Vert_{\Pi} $, while in the second case, we have $
\vert P_k(\overrightarrow{\lambda^0})\vert > 1 \geq \Vert P_k
\Vert_{\Pi}$. Therefore, in both the above cases, we have $
\overrightarrow{\lambda^0} \notin \wp(\Pi) \Rightarrow \wp(\Pi)=
\Pi $ and thus $ \Pi $ is $p$- convex. In general, this lead to
the fact that every p-polyhedron is p-convex. From the definition
of p-convexity and the choice of the polynomial $
P_{\overrightarrow{\lambda^0}} $ $ \forall
\overrightarrow{\lambda^0} \in K^n \setminus K $, the existence of
the above polyhedron implies that we have $ \vert
P_{\overrightarrow{\lambda^0}}(\overrightarrow{\lambda^0}) \vert
>1 $ and $ \Vert P_{\overrightarrow{\lambda}} \Vert_K \leq 1 $.
Moreover, from the continuity condition of $
P_{\overrightarrow{\lambda^0}} $, we see that $ \vert
P_{\overrightarrow{\lambda^0}}(\overrightarrow{\lambda}) \vert >1
$, as for some $ \lambda $, there exists an open neighborhood $
\underline{nbd}( N_{\overrightarrow{\lambda^0}}) $ at $
\overrightarrow{\lambda}= \overrightarrow{\lambda^0} $. Thus,
writing $ P= P_{\overrightarrow{\xi}} $, allowing $ \lambda^0 $
into the range $ P \setminus G ( \subseteq P \setminus K ) $ and
using the compactness of $ P \setminus G $, we have a finite
family of $ \underline{nbd}( N_{\overrightarrow{\lambda^j}}) $,
which corresponds to the polynomials $
P_{\overrightarrow{\lambda^j}} $ such that $\bigcup_{j=1}^n
N_{\overrightarrow{\lambda^j}} \supseteq P \setminus G $. By
setting $ \Pi = \lbrace \overrightarrow{\lambda}\in P \mid \vert
P_{\overrightarrow{\lambda^j}} \vert \leq \Vert
P_{\overrightarrow{\lambda^j}} \Vert_K < 1 \rbrace $, it may
easily be seen that $ K \subseteq \Pi $. Next suppose that
$\overrightarrow{\lambda} \notin G $ with
$\overrightarrow{\lambda} \notin P $, then of course
$\overrightarrow{\lambda} \notin \Pi $, as $\Pi \subseteq P $. On
the other hand, if $\overrightarrow{\lambda} \in P $, then
$\overrightarrow{\lambda} \in P\setminus G $ and thus we have
$\overrightarrow{\lambda} \in N_{\overrightarrow{\lambda^j}} $,
for some $ j $. Hence, it follows that we have $ \vert
P_{\overrightarrow{\lambda^j}} \vert >1 $, whenever
$\overrightarrow{\lambda} \in \Pi $, \textit{viz.} $ \Pi \subseteq G $.
This amounts to say that the polysubset $ \Pi $ can be examined by
the following property. Let $ K $ be a p-convex set such that $ K
\in P_{\overrightarrow{\xi}} $ and $ G $ be an open set with $ K
\subseteq G \subseteq K^n $, then there exists a polyhedron $ \Pi
$ such that $ K \subseteq \Pi \subseteq G $.

In order to analyze the generalized spectrum of a given entropy
function, it requires to recall some basic properties of the abstract
algebraic system. A triple $ (\Re,+,.) $ is said to be a ring
\cite{algebra,algebra1,algebra2}, if the followings hold: (i) $
(\Re,+) $ is abelian, (ii) $a.b \in \Re, \forall a,b \in \Re $,
(iii) $ a . (b. c )= (a . b) . c, \forall a,b,c \in \Re $, (iv) $
a.(b+c)= a.b + a.c $ and $ (a+b).c = a.c + b.c $. Moreover, $
(\Re,+,.) $ is commutative, if $ a.b = b.a $. Subsequently, an
associative ring $ \Re $ becomes an algebra over a field $
\mathcal K $, if $ \Re $ is a vector space over $ \mathcal K $
such that $ \forall a,b \in \Re , \alpha \in \mathcal K $ satisfying 
the following multiplication property $ \alpha (a.b) = (\alpha a).b =
a. ( \alpha b)$. Furthermore, let $ A $ be a linear vector space
over the field $ \mathcal K $ with $ \alpha_1, \alpha_2 \in A $
and $ a_1, a_2 \in \mathcal K $ such that (i) $ a_1 \alpha_1 + a_2
\alpha_2 \in A $ (ii) $ a ( \alpha_1 \alpha_2 ) = (a \alpha_1)
\alpha_2 = \alpha_1 (a \alpha_2 ) $. Then, the algebra $ A $
becomes a morphism, if there exists a map $ \mathcal K \times A
\rightarrow^f A $. For a given ring $ (\Re,+,.) $, we shall take
an account of the fact that there exists an ideal $ I \subseteq
(\Re,+,.), $, if (i) $ a-b \in I; \forall a,b \in I $ and (ii) $
r.a, a.r \in I; \forall a \in I, r \in \Re $. In the case when $
J \neq 0 $ and $ I \subseteq J $, then $ I= J $ which will be
termed as the maximal ideal of the algebra $A$. Further, along
with the other notions of commutative algebra
\cite{algebra,algebra1,algebra2}, we wish to focus our attention
on the homomorphism, which can be thought as a mapping from one
algebraic system to another algebraic system and preserves the
given set theoretic structures. For example, in the case of the
group theory, the map $ \phi $ from a group $ G $ to $ \tilde{G} $
is said to be homomorphism, if $ \forall a,b \in G $ and $
\phi(a),\phi(b) \in \tilde{G}$, we have $ \phi(a.b)= \phi(a)
\phi(b) $. With this reconsideration, we are in a position to
consider the generalized spectrum of arbitrary entropy function,
\textit{viz.} for a finite set of elements, we are in proper position 
to define the complexification map(s). For an even number of elements, 
we shall herewith show that the above consideration leads to
stability properties of the chosen entropy function, \textit{e.g.},
algebraic fronts of the symplectic structure.

Let us consider unital commutative algebra $ A $ with a given unity
element $ e $. For given set of $ a_j \in A $, we may define a
vector as $ \overrightarrow{a}:= (a_1, ..., a_n) $. Subsequently,
if $ \overrightarrow{\lambda}= (\lambda_1, ..., \lambda_n)$ is a
finite collection of the eigenvalues $\lambda_j \in K $, then $ I=
I(\overrightarrow{\lambda})= I(\overrightarrow{\lambda},
\overrightarrow{a})$ is an ideal of $A$ with the following
generators $ (a_j - \lambda_j e); \forall j= 1, ..., n $. For a
given algebra $A$, we may thus define the Krull as $ I=
I(\overrightarrow{\lambda})= \sum_{j=1}^n A (a_j - \lambda_j e) $.
Therefore, the corresponding joint spectrum $ \sigma
(\overrightarrow{a})= \sigma(a_1, ..., a_n) $ is defined by the
set $ \sigma (\overrightarrow{a}):= \lbrace
\overrightarrow{\lambda} \in K^n \vert I(\overrightarrow{\lambda})
\neq A \rbrace $. Notice that the joint spectrum is sometime
termed as the simultaneous and generalized spectrum. For $ n= 1$,
it follows that we have $\overrightarrow{a}= a $ and $
\sigma(\overrightarrow{a})= \sigma(a) $. Further, it is easy to
show that the various properties of the generalized spectrum are
summerized as per the followings: (i) The isomorphism $ I(\overrightarrow{\lambda})
= A $ gives the bijection $ \overrightarrow{\lambda} \in K^n
\setminus \sigma(\overrightarrow{a}) \Leftrightarrow \exists
\lbrace b_i \rbrace_{i=1}^n \in A $ such that $\sum_{j=1}^n (a_j -
\lambda_j e) b_j = e $. (ii) The necessary and sufficient
condition for $ I(\overrightarrow{\lambda}) \neq A $ implies that
$ \overrightarrow{\lambda} \in \sigma(\overrightarrow{a}) $, iff $
\forall \lbrace b_i \rbrace_{i=1}^n \in A$, then there exists
$\sum_{j=1}^n (a_j - \lambda_j e) b_j \notin G_i $; where $ G_i $
denote the invertible group elements of the algebra $ A $. (iii)
If $ \overrightarrow{\lambda} \in \sigma(\overrightarrow{a}) $,
then $ I(\overrightarrow{\lambda}) \neq A $. Thus, the Krull $
I(\overrightarrow{\lambda})$ is contained in the maximal ideal $ M
$, \textit{viz.} $ (a_j - \lambda_j e) \in I(\overrightarrow{\lambda})
\subseteq M $. On the other hand, for some $ j $, if $ (a_j -
\lambda_j e) \in M $, then $ I(\overrightarrow{\lambda}) \subseteq
M \neq A $. Hence, $ \overrightarrow{\lambda} \in
\sigma(\overrightarrow{a}) $, iff there exists a maximal ideal $ M
$ of $ A $ such that we have $ (a_j - \lambda_j e) \in M; \forall
j= 1,..., n $. Consequently, we have the following physical
result, which follows as the corollary of the above properties of
the generalized spectrum: $ \forall \overrightarrow{\lambda} \in
\sigma(\overrightarrow{a}) \Rightarrow \lambda_j \in \sigma(a_j);
j= 1, ..., n $ such that $ \sigma(\overrightarrow{a}) \subseteq
\bigotimes_{j=1}^n \sigma(a_j) $. For given $
\overrightarrow{\lambda} \in \sigma(\overrightarrow{a})$, the
present consideration shows that we have $(a_j - \lambda_j e) \in M
\Rightarrow \lambda_j \in \sigma(a_j) $.

We now exploit the algebraic advantages arising from the
consideration of the complex number field $ C $. Specifically, we
wish to explicate the situation for the spectrum of Sen entropy
function $ F(\overrightarrow{s}) $. For the complexification of 
even dimensional strictly real topological algebras, a number of
interesting cases arise from the character homeomorphism. From the
fact that the complex number field $ C $ is algebraically closed, it
application turns out to be relatively easier to deal with. For
example, there exists a class of matrices and operators, which
have complex eigenvalues. Furthermore, there exist many more
algebraically closed polynomials over the number field $ C $.
Although, we loose the algebraic order relations of the
corresponding real number field $ R $, however, the algebraic
closure property, as one of the main benefits of the field $ C $,
offers us interesting fronts. To illustrate an idea of the power
of the complex number field, let us consider the following matrix
$ \Lambda:= \left (\begin{array}{rr}
    0 & -1  \\
     1 & 0  \\
\end{array} \right) $.
Let $ e_1:= \left (\begin{array}{r}  1 \\ 0 \\ \end{array} \right)
$ and $ e_2:= \left (\begin{array}{r} 0 \\ 1 \\ \end{array}
\right) $ be $ R^2 $ basis. It follows that $ \Lambda .e_1= e_2 $
and $ \Lambda .e_2= -e_1 $ define the corresponding Cauchy Riemann
conditions over the complex number field $C$. Thus, the operator $
\Lambda:C^2 \rightarrow C^2 $ has a well-defined complex
structure. Furthermore, let $ \lambda $ be the eigenvalue of $
\Lambda $, then it follows that we have $ \lambda^2+ 1 =0 $, i.e.
$ \lambda= \pm i \in C $. Therefore, the associated polynomial $
P(\lambda):= \lambda^2+ 1 $ is not closed over $ R $. However, if
possible, let us suppose on the contrary that $ \lambda \in R $
with $ \lambda^2= -1 $ then $ 0 \in R $ and $ (-\lambda)>0
\Rightarrow (-\lambda)(-\lambda)>0 \Rightarrow \lambda^2>0 $.
However, in the case of the real number field, $ \lambda^2= -1<0 $
is a contradiction. We can prove similar contradictions for the
choices $ (-\lambda)<0 $ and $ \lambda= 0 $. Thus, the polynomial
$ P(\lambda) $ possesses root(s) over the number field $ C $,
however, it has no roots over the real number field $ R $.
Consequently, the polynomial $ P(\lambda) $ is algebraically
closed over the complex number field $ C $. This is one of the
main motivation of the subsequent consideration that we may
complexify the spectrum in an even dimensional number field $K^{2n} $.

Motivated from the spectrum, let us examine the role of the
complexification in the corresponding case of Sen entropy
function. In the algebraic setting, an abstract notion of the
complexification can be realized as follows. Let $ A $ be an
algebra over the number field $ \mathcal K$, then there exists a
homomorphism $ \chi $ of $A$ onto $ \mathcal K$ such that its one
dimensional representation corresponds to the character of $A$.
Thus, it follows that $ \chi^{-1}(0)= ker \chi $ is the kernel of
$ \chi $. Subsequently, let $ \Delta= \Delta (A) $ denotes the set
of all possible characters of $ A $. Then, in general case, we
notice further that $ \Delta $ could be empty, in principle. As
per the present consideration, a map $ \chi: A \rightarrow
\mathcal K $ is the character homomorphism, iff there exists a
nontrivial character $ \chi \neq 0 $. It follows further that $
\chi = 0 $ lead to the fact that there exists an element $a_0 \in
A $ such that $ \chi (a_0) = \alpha \neq 0 $. Moreover, if $ \beta
\in \mathcal K $, then $ \chi(\beta \alpha^{-1} a_0) = \beta$
implies that $\chi $ is a surjective map and thus $\chi $ is the
character. Conversely, it is easy to see that, if $\chi $ is non
trivial, then it has a character representation. In general, the
perspective of the complexification may be seen as follows. Let $
A $ be a strictly real topological algebra. For a given
complexification $ \tilde{A} $, the map $ \Lambda: \chi
\rightarrow \tilde{\chi} $ is the homeomorphism such that $ \chi
\in \Delta (A) $ and $ \tilde{ \chi } \in \Delta (\tilde{A}) $.
Similarly, let $ \Lambda_c= \Lambda \vert_{\Delta_c} $ be a
restriction of $ \Lambda $ onto the subspace $ \Delta_c $, then $
\Lambda_c: \Delta_c (A) \rightarrow \Delta_c (\tilde{A}) $ is also
a homeomorphism. Therefore, for a given real valued function $ f $
on $ A $, an extension function $f$ can be represented as $
\tilde{f}(s_1+ s_2)= f(s_1) + i f(s_2), \forall s_1, s_2 \in A $,
which is the standard canonical extension of a real valued function 
$f$ over $C$. It is not difficult to show that the character $
\tilde{ \chi } $ of $ \tilde{A} $ is real valued, if the
corresponding restricted character $ \chi = \tilde{ \chi
}\vert_{A} $ is real. Herewith, we observe that $ \Lambda $ is a
bijection, since $ \tilde{ \chi }(z)= \chi(s_1) + i \chi(s_2),
\forall z = s_1 + i s_2 $; where $ z \in \tilde{A}; s_1, s_2 \in A
$. Moreover, $ \tilde{\chi} $ is continuous, iff $ \chi $ is
continuous. Thus, it is not difficult to show that $ \Lambda_c $
is also a bijection. In the above consideration, there exists a
map $ \tilde{\chi}_{\alpha} \rightarrow \tilde{\chi} $, whenever $
\chi_{\alpha} \rightarrow \chi $ and thus $ \Lambda $ and $
\Lambda_c $ are well-defined homeomorphisms.

Now, let us explicate the above characterization for the joint
spectrum of an entropy function. Although we can generalize the
spectra for the case of non-extremal black holes, for instance
non-extremal $D_1$-$D_5$ and $D_2$-$D_6$-$NS_5$ configurations
\cite{GarousiGhodsiCai, GarousiGhodsiCai1,GarousiGhodsiCai2}.
However, for purpose of uniformity, we shall focus our specific
attention on Sen entropy function of the extremal black holes. Let $
F(\overrightarrow{\lambda}) $ be Sen entropy function with
finitely many electric-magnetic charges $ \lbrace Q_i:= (p_i, q_i)
\in \mathcal M_{2n} \rbrace $, which, from the perspective of
$S$-duality, are considered as symplectic vectors endowed with the
standard symplectic $2$-form $\omega$. For a commutative vector $
\overrightarrow{a} $ in $ K $ and symplectic vector $
\overrightarrow{b} $ in $ \tilde{K} $, the semi-simplectic
spectrum $ \sigma(\overrightarrow{a}, \overrightarrow{b}) $ can be
examined as per the following consideration. For simplicity, let's
consider a pair $(n_1,n_2)$ such that $ n_1 + n_2 = n $, where
$n_2$ is an even integer with $ \overrightarrow{a}= (a_1, ...,
a_{n_1}) $ over a commutative field $ K $. Whilst, the vector $
\overrightarrow{b}= (b_{n_1}, ..., b_{n_1 + n_2}) \in \tilde{K} $
is defined such that there exists an ensemble of pairs $(p_i, q_i)
$, which allows a symplectification over the basis of the algebra
$ \tilde{A} $ and satisfies the following splitting $ \mathcal
A:= A + \tilde{A} $. In this case, we find that the entropy
function $F$ has a semisymplectifiable algebra $ \mathcal A $,
whose semisymplectic spectrum can be expressed as $
\sigma(\overrightarrow{c})= \sigma(\overrightarrow{a},\overrightarrow{b}) $. 
From the perspective of commutative algebra, we notice that the generalized
or joint spectrum of $F$ is defined as per the definition of the convex hull.
Thus, it can be easily shown that the complexification of the
spectrum of $F$ can be accomplished in an algebraically extendable field
of the eigenvalues of the spectrum of $ \mathcal B $ over the
algebra $ \mathcal A:= A + \tilde{A} $.

Before concluding the present section, we wish to illustrate further
consequences of the foregoing spectra. From the perspective of the
thermodynamic and attractor configurations, an unified geometric
interest is associated to the definition of Sen entropy function
and the corresponding attractor entropy of the black hole. For the extremal
black hole configurations, it is worth emphasizing that the results
following from the definition of the Hessian matrix of $F(s)$ is
shown to be important from the perspective of the intrinsic geometry
and commutative algebra. Explicitly, let us focus our attention on
Sen entropy extrimization method, then the above mentioned results
concerning the joint spectra of the Hessian matrix of $F$ offer a
guideline principle towards the stability of the black hole. In particular, 
let us consider the quadratic form of the above entropy function $F$ and 
the associated attractor valued entropy. For the extremal black holes, 
the spectrum pertaining to the entropy function geometry is the product of 
the spectra of the attractor entropy and the associated moduli. In the other 
words, the intrinsic semisymplectic geometry renders to the standard
thermodynamic geometry at the attractor horizon configuration,
when the entropy function $ F $ is evaluated at the extremum
values of the moduli $ \overrightarrow{u} $, gauge fields and
gravity parameters $ \overrightarrow{v} $. The extremization of $
F $ determines the attractor values of $ \overrightarrow{u} $ and
$ \overrightarrow{v} $ in terms of the charges of the black hole.
For given $ \overrightarrow{p} $ and $ \overrightarrow{q} $, the
corresponding generalized attractor equations are given by $
\frac{\partial F(\overrightarrow{s})}{\partial u_s}= 0$ and $
\frac{\partial F(\overrightarrow{s})} {\partial v_i}= 0$. This
yields the critical points $(u_s, v_i)= (u_s^0, v_i^0)$ of the
theory. Subsequently, the entropy of the black hole is evaluated
as the attractor fixed point value of Sen entropy function
$F(\overrightarrow{s})$. Thus, for $ s_i^0= (u_i^0, v_i^0, q_i,
p_i) $, we have the following attractor entropy $
S_{BH}(\overrightarrow{q}, \overrightarrow{p})=
F(\overrightarrow{u}^0, \overrightarrow{v}^0, \overrightarrow{q},
\overrightarrow{p})$. At the attractor fix points $ s_i^0 $,
it follows that the nonzero metric elements of $\mathcal M$ are 
alike from the elements of the standard attractor fixed point geometry.

In order to analyze the spectrum of the attractor flow, let us
define $ r_{ij}:= \frac{\partial^2 S(\overrightarrow{Q})}{\partial
Q^i \partial Q^j} $ as the thermodynamic metric tensor associated
to the charges $Q_i:= (p_i, q_i) $ and thereby consider the
associated matrix $ \mathcal R:=(r_{ij})$. As defined in section
$4$, for given $ b_{ij}:= \frac{\partial^2 F(s)} {\partial s^i
\partial s^j} $ of the entropy function $F$, let us consider the matrix
$ \mathcal B:= (b_{ij}) $. Let $ \lbrace \lambda_i
\rbrace_{i=n_1}^{n_1 + n_2} $ be the eigenvalues of the matrix $
\mathcal R $. As $ n_2 $ is an even integer, thus $ \lbrace
\lambda_i \rbrace_{i=n_1}^{n_1 + n_2} $ form the symplectic
structure under the complexification. For the matrix $ \mathcal B
$, let $ \lbrace \lambda_i \rbrace_{i=1}^{n_1 + n_2} $ be a finite
collection of the eigenvalues of $\mathcal B$, where $ n_2 $ of $
\lbrace \lambda_i \rbrace $ posses the same structures, while $
n_1 $ of the eigenvalues belong to the field $ K $. The present
consideration offers interesting consequences for both the
attractor mechanism and Sen entropy function method. Specifically,
since a given theory has a fixed number of charges, thus the joint
spectra of both the thermodynamic and extended thermodynamic
configurations posses some nonzero complexifiable spectra $
\sigma (b_1,b_2, \ldots, b_{n_2} ) $, which are defined as per the
minimal extension of eigenvalues of the matrices $ \mathcal B
$ and $ \mathcal R $. Therefore, the stability of Sen entropy
function of an extremal black hole can be geometrically examined
by the convexity of eigenvalues of the quadratic function
$\mathcal B$. Algebraically, the corresponding spectrum may be
analyzed as per the definition of the Krull of the algebra of 
eigenvalues of $\mathcal B$. Notice further that the
transformations to the symplectic and real manifolds and to the
associated algebras depend solely on the choice of the pair
$(n_1,n_2)$. Physically, the pair $(n_1,n_2)$ is fixed by the
compactifying manifold. Up to some homomorphism, the spectrum $
\sigma(\overrightarrow{c}) $ may differ from the attractor valued
thermodynamic spectrum $ \sigma(\overrightarrow{b}) $ of Sen
entropy function. It is worth mentioning that there exists a
submersion of the intrinsic metric tensor $g_{s^i
s^j}(\overrightarrow{s}) \vert_{s^i_0} = g_{Q^i
Q^j}(\overrightarrow{Q})$. From the perspective of
complexification of the eigenvalues, we find that both the above
mentioned spectra have the same non-zero joint spectra, up to
a homomorphism. In the next section, we shall illustrate that the
notion of algebraic geometry offers guidelines pertaining to the
stabilization of Sen entropy function.
\section{Calabi-Yau Geometry}\label{six}
In this section, we focus our attention on algebraic properties 
of the two sequences of the embeddings $ \mathcal D_1, \mathcal
D_2 $. As mentioned in the previous section, we shall explicate
here the fact that the stability relations, following from the 
eigenvalues of $ \mathcal B $, yield deformed S-duality transformations.
Pertaining to the mentioned issues in sections $4$ and $5$, we wish to 
illustrate the case of the Calabi-Yau moduli. For a set of given scalar 
fields, we present the associated algebraic geometric properties from
the perspective of $A$ and $B$ modules. As mentioned in section
$1$, let us begin our analysis by reconsidering the notions from the
Refs. \cite{HitchinLiYau,HitchinLiYau1,Calabi1975}, namely, a $n$-dimensional
complex manifold $ X $ endowed with local coordinate charts $
\lbrace z^i, \overline{z}^j \rbrace $. Let $ B= B_{i_1,
i_2,\ldots, i_p;j_1, j_2,\ldots, j_q} d z^{i_1} \wedge d z^{i_2}
\wedge \ldots d z^{i_p} \wedge d \overline{z}^{j_1} \wedge d
\overline{z}^{j_2} \wedge \ldots d \overline{z}^{j_q} $ be a $
(p,q) $-form on $X$. Then, the exterior derivative operator $
\partial $ is expressed by $
\partial B= \frac{\partial }{\partial z^k} B_{i_1, i_2,\ldots,
i_p;j_1, j_2,\ldots, j_q}d z^k \wedge d  z^{i_1} \wedge d  z^{i_2}
\wedge \ldots d  z^{i_p} \wedge d \overline{z}^{j_1} \wedge d
\overline{z}^{j_2} \wedge \ldots d \overline{z}^{j_q} $.
Furthermore, a similar expression follows for the operator $
\overline{\partial} B $. Subsequently, it is easy to see that $
\partial^2= 0= \overline{\partial}^2 $. Under the consideration
of Dolbeault cohomology, the Hodge theorem amounts to the fact
that each cohomology class $ \mathcal H^{p,q}(X) $ contains unique
harmonic form $B_k$ such that the Laplacian operator decomposes as
per the following $ \Delta_{\overline{\partial}}=
\overline{\partial \partial^{\star}}+ \overline{\partial}^{\star}
\overline{\partial} $, where $ \Delta_{\overline{\partial}} B_h= 0
$. In the case of K\"{a}hler manifold, the above Laplacian leads
to the standard notion that we have $ \Delta= 2
\Delta_{\overline{\partial}}= 2 \Delta_{\partial} $. Thus, the
harmonic forms associated with the derivative operators $\partial
$ and $ \overline{\partial} $ imply an isomorphism between the
vector spaces $ \mathcal H^{p,q}_{\partial}(X) $ and $ \mathcal
H^{p,q}_{ \overline{\partial}}(X) $. For any $ (p,q) $-harmonic
form $ B_{p,q} $ on $X$, such an identification implies the
existence of de-Rham cohomology. For $(p, q)$-form, it turns
out that $ B_{p,q} $ satisfy the following Dolbeault sum $ B_p=
B_{p,0}+ B_{p-1,1}+ \ldots+ B_{0,p}$. For a given $p$-form $ B_p
$, an application of the operator $ \Delta $ on $ B_{p} $ shows
that we have $ \Delta B_p= 0 $. Thus, $ \Delta_{\overline{
\partial}} B_{p_1,p_2} $ becomes a $ (p_1, p_2) $-form, whenever $
\Delta_{\overline{\partial}} $ preserves the degree of $
B_{p_1,p_2} $. It is easy to see that the principle of mathematical 
induction on the index $ i $ gives $ \Delta_{\overline{\partial}}B_{p-i,i}=
0, \forall i $. Thus, the vector space pertaining to the harmonic
de-Rham $ p $-form decomposes as the direct sum of the harmonic
Dolbeault $ (p_1, p_2) $-form, where $ p_1+ p_2= p $.
Consequently, in the case of the K\"{a}hler manifold, the harmonic
forms represent a set of cohomology classes, which uniquely
reduces to the following decomposition $ \mathcal H^p(X)= \mathcal
H^{p,0}(X)\oplus \mathcal H^{p-1,1}(X)\oplus \ldots \oplus
\mathcal H^{0,p}(X) $. In this case, the betti numbers are given
by $ b^p= h^{p,0}+ h^{p-1,1}+\ldots+ h^{0,p} $, where $ h^{i,j}=
dim(\mathcal H^{i,j}(X)) $ are the Hodge numbers of $ X $.

The manifold $X$ becomes a K\"{a}hler manifold, if there exists a
complex valued function $ K(z,\overline{z}) $ such that the
underlying metric tensor is defined by $ g_{i\overline{j}}=
\frac{\partial^2 K(z,\overline{z})}{\partial z \partial
\overline{z}} $. As mentioned in section $2$, there exists a
non-degenerate closed 2-form $ \omega= 2i \partial \overline{
\partial} K(z,\overline{z}) $ such that $ d \omega= 0 $ for any $K$.
On the other hand, it is well known that the topological
invariance of K\"{a}hler manifold imposes a pair of
restrictions on the associated Hodge numbers, \textit{viz.} $ h^{p,q}=
h^{q,p} $ and $ h^{p,q}= h^{n-p,n-q} $. For a set of harmonic
$(q,p)$ forms on the given K\"{a}hler manifold, the first
restriction corresponds to a mapping of the operator $ \partial $
to the operator $ \overline{\partial} $. Thus, the correspondence
$ B \rightarrow \overline{B} $ results into an invertible map
between the underlying $ \partial $ and $ \overline{\partial} $ cohomologies.
Further, the second restriction could be realized as the map $ (A,
B) \rightarrow \int_X A \wedge B $ from the space $ \mathcal
H^{p,q} \oplus \mathcal H^{n-p, n-q} $ into $ C $. As the above
map is non-degenerate, thus $ \mathcal H^{p,q} $ and $ \mathcal
H^{n-p, n-q} $ can be viewed as a pair of dual vector spaces with
the same dimension $ b^p= b^{n-p} $. In correspondence with the
associated de-Rham cohomology, the identification of $ p $-form cohomology
class with $ (n- p) $ cycle homology class leads to the standard
Poincar\'{e} duality. For a given harmonic class $ \Sigma $, the
perspective of homology classes follows from the consideration of
$ (n-p) $ cycle on $ \mathcal H_{n-p}$. In the limit $ p
\rightarrow 0$, there exists a class of delta function $ \delta(\Sigma) $
localizations on $ p $-dimensional hypersurfaces. Such a
consideration examines the limiting parametric nature of $ p
$-forms. Considering the fact that $ \delta(\Sigma) $ is defined
as an integral over $p$-dimensional submanifold, Poincar\'{e}
duality of $ \Sigma $ provides the underlying cohomology classes.
The Hodge integers of K\"{a}hler manifolds remain symmetric under
both the horizontal and vertical reflections.

In order to examine the case of Calabi-Yau, let us consider $n$
complex dimensional k\"{a}hler manifold endowed with a
covariantly constant holomorphic $n$-form $ \Omega $ such that
there exists a Riemannian manifold with $SU(n)$ holonomy. In
particular, let us consider the case of $ n $-torus $ T^n=
\prod_{i=1}^n (S^1)^i $. Let $ \lbrace X_i \rbrace_{i=1}^n $ be a
set of generating vector fields which preserve complex structure $
J $ and $ \lbrace J X_i \rbrace_{i=1}^n $ as the corresponding
trivialization of the tangent bundle $ T \mathcal M $. Given $
Z_i:= X_i- J X_i, \forall i= 1,2, \ldots, n $, we can define $n$
commuting holomorphic vector fields $ \lbrace Z_i \rbrace_{i=1}^n
$ such that there exists a subset $ \mathcal M_0 \subset \mathcal
M $, which is biholomorphically equivalent to an open neighborhood of
the set $ (S^1)^n \subset (C^{\star})^n $. Thus, the definition of
the holomorphic vector fields follows from the trivialization of
the canonical tangent bundle $ T \mathcal M $. Moreover, one can
choose $ \mathcal M_0 $ to be fibred over a ball $ B_0 \in B $
such that $\mathcal M_0 $ is homotopically equivalent to an orbit
of $ T^n $. It follows from the notion of the Lagrangian manifolds 
that the K\"ahler form $ \omega $ restricts to the zeroes of 
the orbits, \textit{e.g.}, $ \omega $ has a trivial cohomology
class. Hereby, for $\theta \in \Omega^1(\mathcal M_0 ) $, there
exists a $(1, 1)$ form $\omega = d \theta = d(\theta^{1,0} +
\theta^{0,1} ) $. As the higher Dolbeault cohomology groups of the
product of open sets vanishes in $ C^{\star} $, thus there exists a function 
$f$ such that $ \overline{\partial} f = \theta^{0,1} $. From the
definition of K\"ahler potential $ K = Im f $, we have $ \omega= d(\partial
\overline{f} + \overline{\partial}f)= 2i \partial
\overline{\partial} K $. In order to obtain $ T^n $-invariance of
the K\"ahler potential $K$, we need to take an average over the compact 
group $ T^n $. By defining the map $(z_1 ,\ldots, z_n ) \mapsto
(e^{z_1},\ldots ,e^{z_n} )\in (C^{\star})^n $, one finds the
following Kh\"aler potential $ K(z_1,\ldots, z_n) = K(z_1 +
\overline{z_1}, \ldots, z_n + \overline{z_n})$. Thus, the Kh\"aler
metric can be considered as the function of the real parts of 
holomorphic coordinates. In this concern, the work of Calabi
\cite{Calabi1975} offers an interesting front for the construction
of nonhomogeneous Einstein metrics.

Perspective situations arise further, when the underlying Ricci
tensor vanishes identically. In this case, there exists a
covariantly constant holomorphic $n$-form $ dz_1 \wedge \ldots
\wedge dz_n $, iff the Kh\"aler potential $ K $ satisfies the real
Amp\`ere-Monge equations, \textit{viz.} we have 
$ det(\frac{\partial^2 K}{\partial x_i \partial x_j})=$ constant. 
With the help of the relative coordinates $ \lbrace x_i, y_i \rbrace $, 
the metric can be expressed as $ g= \sum_{ i,j} \frac{\partial^2 K} {\partial x_i
\partial x_j}(dx_i dx_j + dy_i dy_j)$. Considering the fact that
$ T^n $ acts isometrically on $\mathcal M $, one finds that there
exists a non-trivial quotient metric on the base space $ B $. From 
the perspective of the moduli space geometry, such a metric can be
defined as the quotient metric on special Lagrangian submanifold
$\mathcal M $. For a set of $ T^n $-invariant coordinates $
\lbrace x_i \rbrace $, the metric tensor locally reduces into the
following form $ g= \sum_{i,j}\frac{\partial^2 K}{\partial x_i
\partial x_j} dy_i dy_j $. Explicitly, the consideration of the local
coordinates $ z_i = x_j + iy_j $ implies that the orbit of $ T^n $
can be viewed as the point in $B$ such that $(x_1 , . . . , x_n )=
(e^{iy_1},\ldots ,e^{iy_n}) $. Thus, the set $\{ \frac{dy_1}{2
\pi},\ldots, \frac{dy_n}{2 \pi} \}$ characterizes forms whose
de-Rham cohomology classes form the integral basis of the first
cohomology class. For each fibre $\mathcal M_0 \rightarrow B_0 $,
this follows from the fact that $ Jdx_i = dy_i $, where $ J $ is
orthogonal matrix. Moreover, the metric on each of the torus can
be reduced to the flat metric $ g= \sum_{i,j}\frac{\partial^2
K}{\partial x_i \partial x_j} dy_i dy_j $. Herewith, there exists
a map from the base space $ B_0 $ to the moduli space of real flat
$n$-tori, which can be identified as the space of $ n \times n $
positive definite matrices with $ \frac{GL(n, R)^{+}}{SO(n)} $.
Up to finitely many modulo actions of $ GL(n, Z) $, the subsequent
analysis is devoted to examine the fact whether the map 
$ x \rightarrow \frac{\partial^2 K}{\partial x_i \partial
x_j} $ offers a globally well-defined integral basis $ \{
\frac{dy_1}{2 \pi},\ldots, \frac{dy_n}{2 \pi} \}$ for $
H^1(\mathcal M, R) $.

From the perspective of compactification, the above notion
demonstrates a lower dimensional realization of string theory. 
Physically, it is worth mentioning that a local Calabi-Yau manifold 
$X$ can be defined as the equation $ s_1 s_2 = H(s_3, s_4) $ in $C^4$
such that there exists a well-defined holomorphic $(0,3)$ form
$ \gamma = \frac{ds_1}{s_1}\wedge s_3 \wedge s_4 $. Here, the manifold $X$
can be regarded as $ C^{\star} $ fibration of the fiber $ s_1 s_2 =$
constant over the $ s_3 s_4 $ plane. Thus, the corresponding $3$-cycles 
on $X$ reduces to the associated $ 1 $-cycles on the Riemann Surface
$\Sigma : 0 = H(s_3, s_4) $. In fact, the periods of $ \gamma $ on
$ X $ descend down to the periods of the meromorphic $ 1 $-form $
\Lambda $ on $\Sigma $ such that $ \int_{3-cycles} \gamma=
\int_{1-cycles} \Lambda $, where $ \Lambda:= s_3 ds_4 $. For arbitrary
$ g $ genus Riemann surface $ \Sigma $, there exists $ 2g $ compact $ 1
$-cycles such that the symplective basis $\lbrace A^i, B_j \rbrace
$ satisfy $ A^i \cap B_j = \delta^i_j ; i,j= 1,2, ..., g $. For a
general noncompact Riemann surface, we may work with an associated
set of general basis, which are defined as the intersections $ A^i
\cap B_j = n^i_j $, where $ n^i_j $ are integrals. To proceed
further, let us define $ q^i = \int_{A^i} \Lambda , p_i =
\int_{B_i} \Lambda $ and thus the set $ \lbrace q^i \rbrace $ can be
considered as the normalizable moduli of $ X $. In the above case,
for a given noncompact CY, we observe that the function $ H(s_3,s_4) $
depends on a set of non-normalizable complex structure moduli $
\lbrace t^{\alpha} \rbrace $ such that there exists compact $ 3
$-cycles $ C_{\alpha} \in H_3(X) $ and $ 1 $-cycles on $ \Sigma $,
where $ t^{\alpha}= \int_{C_{\alpha}} \Lambda $. Herewith, it
follows that the moduli space metric remains non-normalizable in
the directions for which the corresponding homology dual cycles $
C^{\alpha} $ are non-compact. Consequently, the parameters
$\lbrace t^{\alpha} \rbrace $ of the model are not themselves 
the physical moduli. Nevertheless, the present consideration implies a
set of deformed S-duality transformations $ \tilde{p}_i= A_i^j p_j
+ B_{ij} q^j + F_{i \alpha} t^{\alpha}$ and $ \tilde{q}^i= C^{ij}
p_j + D^i_j q^j + F^i_{\alpha} t^{\alpha}$, where $ \lbrace
t^{\alpha} \rbrace $ can be interpreted as a set of monodromy
invariant parameters. In this case, notice further that we have
defined the most general extension maps $ q^i \rightarrow
\tilde{q}^i, p_i \rightarrow \tilde{p}_i $ such that the monodromy
group preserves the symplectic form $ dq^i \wedge dp_i $ with $ \left
(\begin{array}{rr} A & B \\  C & D \\ \end{array} \right) \in
Sp(2g,Z) $. In the foregoing case, it is easy to show that the
basis $ \lbrace q^i, p_j \rbrace $ transforms in an unconventional
way, while the deformed period matrix $ \tau_{ij}=
\frac{\partial}{\partial q^j} p_i $ transforms in the standard
way. Explicitly, this leads to the following period map: $ \tau
\rightarrow \tilde{\tau} = \frac{A \tau + B}{C\tau + D} $. Now, it
is natural to think of $ H^3(X,Z) $ as the classical phase space
pertaining to the symplectic form $ dq^i \wedge dp_i $. In order
to quantize $ X $, we may promote $\{ q^i, p_j \}$ to their
canonical conjugate operators satisfying the following canonical
commutation relations $ [q^i, p_j]= \hbar \delta^i_j $.
Furthermore, the above method of quantization remains compatible 
on a general CY, including both the real and complex polarizations. 
This follows from the fact that $ (0,3) $ form $ \gamma $
lives in the complexification $ H^3(X,C) = C \otimes H^3(X,R)$. As
in the case of the compact CY, the period integrals over $ A_I $
and $ B^J $-cycles over $ CY(3) $ can be expressed in terms of the
moduli, for a given $(0,3)$-form $ \gamma $. For example, the
background independent theory \cite{Witten,Witten1} offers further
algebraic issues. For the case of the $B$-model topological string theory 
and almost modular forms, Ref. \cite{topstrings} offers Gromov-Witten 
invariants of a class of orbifold theories and the corresponding quantum mechanical 
phase space properties, as the quantum geometry of $H^3(X)$, where $X$ 
is a Calabi-Yau threefold. In the present consideration, we may hereby 
define the moduli $ s_I $ and $ \tilde{s}_I $ as the generalized period 
integrals over the cycles $ A_I $ and $ B^I $. For a given $ \gamma $, 
the quantization may be analyzed by the following integrals: $ \int_{A^I} \gamma = s^I $ 
and $ \int_{B_I} \gamma = \tilde{s}_I $. Moreover, for cases $ D > 4 $, 
we notice the existence of a larger sequence $ \lbrace u_i \rbrace_{i=1}^{2k} $, 
instead of a short sequence: $ \lbrace u_i \rbrace_{i=1}^{2m} $, where $ k>m $. 
Thus, apart from the above extension of $ \mathcal M_{2m} $ to $
\mathcal M_{2k} $, there is the complete freedom to define the notion
of semisymplectic geometries and thus an interesting deformation
property of the underlying algebraic geometry.

In order to interrelate the properties of the underlying embeddings
and the spectra of Sen entropy function, we may proceed as follow. Let
$ H^{ij}(X) $ be the local vector spaces of the Calabi-Yau manifold $
X $ and $h^{ij}$ be the dimension of the corresponding representations. 
Considering the fact that the Hodge integers $\lbrace h^{ij} \rbrace $ 
form diamond structure with the symmetries $h^{ij}= h^{ji} $ and 
$ h^{i,j}= h^{n-i, n-j} $, it follows that the Hodge integers form 
standard $Z$-module structures and thus their fractions form the
perspective of $Q$-modules. Let the underlying characteristic
polynomials be defined as $ f^{ij}= a_0+ a_1 x+ \ldots + a_n x^n=
0 $, then the root of $f^{ij}$ can be expressed by the set $ \lbrace r_{\alpha}
\rbrace_{\alpha= 1}^n $. In this case, the fact that $ h^{ij} \in
Z; \forall i,j= 1,2, \ldots , n $ allows us to define $
\zeta_{ij}= \lbrace r_1, r_2, \ldots, r_{h^{ij}} \rbrace $, $
\zeta= \bigcup_{i,j=1}^n \zeta_{ij} $ and $ T= \bigcap_{i,j= 1}^n
\zeta_{ij} $. Subsequently, let $ J= \lbrace x \ \vert \ f(x)= 0,
\forall x \in X \rbrace $. Then, we observe that $ \zeta_{ij}= J
$, as $ f(x)= 0, \forall x= r_{\alpha}$, where $\alpha= 1,2,
\ldots h^{ij} $. Moreover, the element $ T $ is the maximal ideal
generated by the set $ \zeta_{ij} $. From the theory of $A$ and
$B$ modules, let $M$ be $A$-module and $N$ be $B$-module. Then,
there exists the following duality map $ \delta: M \rightarrow N
$. For the case of the free modules, we see that the above duality
map gives rise to the standard Poincar\'{e} type pairing
between the spaces $ H^{ij}(X) $ and $ H_{ij}(X) $. Due to the
duality equivalence of $A$ and $B$-modules, we see for given
cohomology and homology sequences that there exists a set of
mappings $ \lbrace t_1, t_2 \rbrace $ such that $ t_1: M
\rightarrow H^{ij}(X) $ and $ t_2: N \rightarrow H_{ij}(X) $, and
vice-versa. In other words, the maps $ t_1 $ and $ t_2 $ are
well-defined up to a duality transformation. From the viewpoint of
Poincar\'{e} duality lemma, it follows that $ \mathcal D: H^{ij}
\rightarrow H_{ij} $ and thus we have an existence of the
following sequences: $ M \rightarrow_{t_1} H^{ij} \rightarrow_{
\mathcal D} H_{ij} \rightarrow_{t_2^{-1}} N \rightarrow_{
\delta^{-1}} M $ and $ N \rightarrow_{t_2} H_{ij} \rightarrow_{
\mathcal D^{-1}} H^{ij} \rightarrow_{t_1^{-1}} M \rightarrow_{
\delta} N $. Hence, it is not difficult to show that there exists
a well-defined composition of $ \lbrace u_1, u_2 \rbrace $, \textit{viz.}
the maps $u_1: M \rightarrow H_{ij}(X) $ and $ u_2: M \rightarrow
H^{ij}(X) $ are well-defined under the composition. Moreover, from
the perspective of $A$ and $B$ module theory, we find that $
H^{ij} $ and $ H_{ij} $ are encoded by the maps $ \lbrace t_i
\rbrace $ and $ \lbrace u_i \rbrace $. Herewith, it follows that
we have (i) $ M \rightarrow_{t_1} H^{ij} \rightarrow_{\mathcal D}
H_{ij} \rightarrow_{t_2^{-1}} N \rightarrow_{\delta^{-1}} M $ and
(ii) $ N \rightarrow_{t_2} H_{ij} \rightarrow_{\mathcal D^{-1}}
H^{ij} \rightarrow_{t_1^{-1}} M \rightarrow_{\delta} N $. For the
above sequences (i) and (ii), we find that the corresponding block
diagram commutes, iff the maps $\{t_i\}$ satisfy $ t_1 \circ
\mathcal D= \delta \circ t_2 $.

A relation between the rational points on $A$-module and the
corresponding complex points on the associated K\"{a}hler manifold $ X $ may
be defined as follow. Let $ M $ be $A$-module generated by the
standard fractional transformations, then the cohomology group of
$ X $ can be described by the map $ t_1: M \rightarrow H^{ij}(X)
$. It is interesting to notice $ \forall z \in C^n $ that the
K\"{a}hler potential $ K(z,\overline{z}) $ defines the characteristic
function of a given K\"{a}hler manifold $ X $. Thus, the minimal
polynomial $ f( s^1, s^2, \ldots, s^m )$ can be defined as a
function in $m $ rational's $ s_{\alpha}:= p_{\alpha}/q_{\alpha} $
such that $ q_{\alpha} \neq 0 $ and $ p_{\alpha}, q_{\alpha} \in Z
$. Let $ \rho_{\alpha}(s^i) $ be the corresponding minimal monic
polynomial such that the roots of $ f(s^i)= \prod_j \rho_{\alpha}^{k_j}(s^i) $
are defined as the set $ \lbrace r_l \rbrace $. Then, for given free
$A$-module, we find that the rational basis functions satisfy $
\lbrace s^1, \ldots s^m \rbrace \rightarrow_{t_1} \lbrace e^1,
\ldots, e^n; \overline{e}^1 \ldots \overline{e}^n \rbrace
\rightarrow_d \lbrace \tilde{e}^1 \ldots \tilde{e}^{h^{i,j}}
\rbrace $. Subsequently, we observe that the composition of $ t_1
$ and $ d $ is well-defined, whenever $ m= 2n $. For a given $M$,
the dimensional equivalence property $dim(M)= h^{i,j}$ implies
that $t_1$ is an isomorphism. Furthermore, we can map the rational
points on $ B $-module to the homology sequence of $X$. Notice
further that the prolongment $ Q \rightarrow R $ lifts $ s^i \in R
$ and thus it describes classical properties of the basis set $
\lbrace s^i \rbrace $. Interestingly, such a prolongment plays an
important role in revealing the limiting relationships between the
classical and quantum moduli spaces.  An exact analysis of the
above question is left open for a future investigation.

In order to make an equivalence between the algebraic and
geometric stability properties of the entropy function of an
extremal black hole, we may ask the following questions. From the
perspective of algebra, the unit circle $ x^2+ y^2= 1 $ can be
examined by considering the Euclidean ring $ k(x,y) \setminus (
x^2+ y^2- 1) $, where $ ( x^2+ y^2- 1) $ is the prime ideal
generated by the unit circle. Such an introduction motivates the
study of algebraic geometric properties of black brane entropy
functions. What follows next that we wish to consider a set of
rings, which are associated with the stability of a given entropy
function. Let us consider the Taylor series expansion of the
entropy function as a finite polynomial in $\{s_i\}$. Then, the
attractor stability of the black hole can be examined by the
algebraic properties of finite polynomials. Notice further that 
an entropy function could lead to a non-polynomial expansion, as well.
For a single variable complex valued entropy function $f(z)$, such
an expansion could lead to the following (in)finite Lorentz series
$f(z)= \ldots + \frac{a_{-2}}{z^2}+ \frac{a_{-1}}{z}+ a_0+ a_1 z+
a_2 z^2+ \ldots, \forall z:= x+ i y \in C $, depending on $a_{i}$,
where $i \in \tilde{\Lambda} \subset Z$ or $i \in Z$. Up to an algebraic
factor, the above cases can be explored polynomially in $z \in C$,
whenever one can normalize the Lorentz series expansion of $f(z)$.
To be precise, let us consider an ideal $(f(z))$ and define the
corresponding generalized Euclidean ring as $ \varpi_0:= f(x,
y)\setminus ( f(x,y) ) $. On the other hand, let the underlying
geometry be defined by the Poincar\'{e} like element $ ds^2:= dz d
\overline{z}/ f(z) $. Inductively, we may ask the following question: 
What are the morphisms that $ \varpi_0 $ preserves? In order to extend
the case for the generalized real K\"{a}hler potential $
K(\overrightarrow{u}, \overrightarrow{v}) $, let $u_i$ be the near
horizon values of the real scalar fields and $v_i$ be the near
horizon spacetime parameters pertaining to $AdS_2 \times
S^{D-2}$ near horizon geometry of the extremal black hole. Let $
g_{ij}=\partial_i
\partial_j K(\overrightarrow{u}, \overrightarrow{v}) $ be the
metric tensor with the corresponding line element $ ds^2= g_{ij} d
\phi^i d \phi^j $, where $\phi^i= (\overrightarrow{u},
\overrightarrow{v})$. Now, define $ \varpi = K(\phi^i) \setminus
(K(\phi^i)) $, where $ (K(\phi^i)) $ is a polynomial ideal
generating the equations of the motion, up to an algebraic factor.
Then, the present analysis anticipates an interesting avenue to examine 
the associated algebraic and geometric structures of the map $ \varpi $ 
and its possible variants. Such an issue pertains to both the attractor 
stability of the black hole and Sen entropy function.

For a given entropy function $ F(\overrightarrow{p},
\overrightarrow{q},\overrightarrow{u}, \overrightarrow{v}) $, let
the metric tensor of the interest be defined as  $ g_{ab}=
-\partial_a \partial_b F(\overrightarrow{s}) $, where $
\overrightarrow{s}:= (\overrightarrow{p}, \overrightarrow{q},
\overrightarrow{u}, \overrightarrow{v}) \in \mathcal M_{2n}
\otimes \mathcal M_{u} \otimes \mathcal M_{v} $. As mentioned
before, let the manifold $ \mathcal M_{2n} $ be symplectic, and
$\mathcal M_{u}$ and $ \mathcal M_{v} $ be the two Riemannian manifolds. 
Thus, the present consideration opens an avenue to analyze the algebraic 
properties of the quotient $ \varpi_1 = F(\overrightarrow{s})\setminus
(F(\overrightarrow{s})) $. In this case, we may preserve the
symplectic structure with $ [ p_i,q_j ]= \hbar \delta_{ij}, \
\forall (\overrightarrow{p}, \overrightarrow{q}) \in ( \mathcal
M_{2n},\omega ) $, where $ \omega $ denotes the corresponding
non-degenerate closed two form. For a given map $ F: \mathcal
M_{2n} \otimes \mathcal M_u \otimes \mathcal M_v \rightarrow R$,
there exists a set of embeddings $ \mathcal M_{2n}
\hookrightarrow_{\mathcal D_1} \mathcal M_{2n} \otimes \mathcal
M_u \hookrightarrow_{\mathcal D_2} \mathcal M_{2n} \otimes
\mathcal M_u \otimes \mathcal M_v $. Herewith, we observe that the
map $ \mathcal D_1 $ is a symplectic embedding, however $ \mathcal
D_2 $ is a deformed symplectic embedding. For given manifolds $
\{ \mathcal M_{2n}, \mathcal M_u, \mathcal M_v \}$, such a
consideration indicates the existence of an unidirectional
composition $ \mathcal D_1 \circ \mathcal D_2$. In general, it
follows further that the other composition $ \mathcal D_2 \circ
\mathcal D_1 $ may as well be ill-defined. Thus, an entropy
function can geometrically be defined as the set of embeddings $
\lbrace \mathcal D_1, \mathcal D_2 \rbrace $ on a given semisymplectic
manifold $ \mathcal M $. From the perspective of the quotient map
$\varpi_1 $, it would be interesting to analyze the algebraic
properties of the above embeddings $ \mathcal D_1, \mathcal D_2 $.
In order to study the nature of the underlying moduli space
compactification, let us consider the restriction $ (
\overrightarrow{p}, \overrightarrow{q})= ( \overrightarrow{p_0},
\overrightarrow{q_0}) $. From the perspective of Sen entropy
function and generalized attractor flow properties, we may ask the
following question. What are the algebraic properties of the maps
$ \varpi_1 $ and $ \varpi $ under the identification $ \tilde{ K}:
\tilde{\mathcal M} \rightarrow R $ satisfying the local structure $
\partial_i \partial_j F(\overrightarrow{s})\vert_{( \overrightarrow{p_0},
\overrightarrow{q_0})}= \partial_i \partial_j \tilde{
K}(\overrightarrow{u},\overrightarrow{v}), \forall
(\overrightarrow{u},\overrightarrow{v}) \in \tilde{\mathcal M}:=
\mathcal M_u \otimes \mathcal M_v $? On the other hand, let us
consider the attractor flow data as the restrictions: $ u_i
\rightarrow u_i^0 $ and $ v_i \rightarrow v_i^0 , \forall i= 1,2,
\ldots, n $. Then, such a consideration offer a platform to
examine an existence of the submersion of $ \mathcal M$ to a
symplectic manifold $\mathcal M_{2n}$. As mentioned before, for
given $b_{ij}$, we may consider the associated line element as $
ds^2:= r_{ij}dQ^i dQ^j $ with the Riemannian metric $ r_{ij}:=
\partial_i \partial_j S(\overrightarrow{Q}) $ and thereby examine
the existence of the natural complex structure $ J $ such that
there exists a non degenerate closed two form $ \omega $ and thus
a symplectic manifold $( \mathcal M_{2n}, \omega )$. Finally, the
properties of $ \varrho:= S(\overrightarrow{Q}) \backslash
(S(\overrightarrow{Q})) $ could reveal underlying algebraic issues
pertaining to the deformed string dualities. Moreover, such a
consideration opens up a set of avenues to examine the algebraic
properties of $ \varpi_1 \vert_{(u_i^0, v_i^0)} $ and $ \varrho $.
For given attractor values of the moduli fields and electric
magnetic charges $\{u_i^0, v_i^0, \overrightarrow{q_0},
\overrightarrow{p_0}\}$ of the black hole, it would be important
to explore the morphism properties of the maps $ \varpi_1 $,
$ \varpi $ and $ \varrho $.
\section{Conclusion and Outlook}\label{seven}
In this note, we have analyzed geometric and algebraic properties
of the Hessian of Sen entropy function, associated with the 
computation of attractor entropy. For a given extremal black hole 
in string theory, we have offered explicit construction for the
generalized symplectic metric, generalized symplectomorphism,
generalized symplectic atlas, generalized symplectic stability,
generalized Legendre transformation, semi-product, strong
stability conditions and thereby illustrated algebraic underlying properties
of the spectrum. In particular, the above analysis of the spectrum
and convexity has been realized from Krull of the semisymplectifiable 
algebra $ \mathcal A $, which arises as the minimally extended (sub)field 
of the underlying eigenvalues of the Hassian quadratic function $ \mathcal B $, 
defining the attractor stability of the underlying black hole configuration,
as per the notion of a semisymplectic manifold $ \mathcal M $. We find
that the semisymplectic geometry, when considered as the minimally
extended commutative algebraic system, describes the algebraic
geometric properties of the black hole entropy functions with
various higher curvature corrections of arbitrary $ D $
dimensional black hole spacetime. Geometrically, the above
analysis reduces to the symplectic geometry with a set of given
generalized complex structures, which can be defined as the
composition of the two series of the embeddings: $ \mathcal D_1 $
and $ \mathcal D_2 $. From the perspective of the extremal black
hole configurations, our algebraic geometric study explains the
attractor stability of $ AdS_2 \times S^{D-2} $ near horizon
geometry and thus it contains possible effects of the scalar
fields, gauge fields and arbitrary higher derivative covariant
gravity. It turns out that the Riemannian geometric viewpoints of
the underlying entropy function renders to the standard thermodynamic 
geometry, at the attractor fixed point values of a given entropy function. 
Both the above two geometries possess the same set of nonzero complexifiable 
generalized spectra, which we have shown to be in the accordance with 
standard attractor flow properties. Thus, we find that the embedding 
perspective of Sen entropy function of a given (extremal) black hole 
configuration encodes the algebraic geometric characteristic features, 
as a generalized attractor.

For the shake of mathematical simplicity, we have chosen the entropy 
function $F(\overrightarrow{s})$ to be a real map. However, for the 
case of the complex effective Lagrangian theories, our construction 
allows a set of geometric transitions, which takes an account of the
transition from a given real manifold to the corresponding complex manifold. 
Thus, our analysis further extends for the complex valued entropy functions. 
In this sense, the non-holomorphic corrections to the prepotential 
appear analogously as the corresponding fluctuations of the underlying
statistical ensemble. In particular, one can easily construct a
complex map $ F: \mathcal M \longrightarrow C $ such that the
generalized complex or complex semisymplectic structure $ b^2 $
remains compatible with deformed Cauchy-Riemann conditions on 
the connection functions $ \lbrace s^{-1}_i\circ s_j \rbrace $.
Indeed, the corresponding higher derivative corrections can be
examined with such a complexified consideration. For a given
generalized prepotential, when the above method is applied to the
CY geometries, we find a set of deformed S-duality transformations
accompanying an inclusion of the monodromy invariant parameters, and
thus an internal compatibility with the CY quantization. Refs.
\cite{WittenStrominger,WittenStrominger1} offer a quantization of
such local space from the perspective of the complex polarization
method on the space $H^3(\mathcal M,C)$ and the corresponding
reality equivalences on $ H^3(\mathcal M,R) $. For the case of the
semisymplectic transformations, we have illustrated that the concerned 
extensions can be defined as the composition of finitely many mappings,
\textit{e.g.}, an identification map $ p_i \mapsto \tilde{p}_i = p_i +u_i +
v_i $. Thus, the present analysis offers an explicit realization
of shifted S-duality transformations, which describe dynamical behavior
of the black hole system away from the attractor fixed point(s).

Moreover, from the perspective of the convexity of the spectrum of
an extremal black hole entropy function, we have shown that the
corresponding algebraic nature emerges from the minimally extended
subfield which contains eigenvalues of the Hessian quadratic
function $ \mathcal B $. For a given semisymplectic manifold $
\mathcal M $, the above consideration leads to a deformation of
the base space symplectic manifold $ \mathcal M_{2n} $ by the
moduli space manifold $ \mathcal M_\phi $ and the gravity sector
manifold $ \mathcal M_k $. Explicitly, our analysis shows that
such generalized spectra can be investigated from the perspective
of commutative algebra, by defining the associated convex hulls.
Herewith, we have established that the complexification can be
realized in the minimally extended subfield of the eigenvalues of
$ \mathcal B $, which at the attractor fixed point(s), reduces to
the standard thermodynamic like configuration and the corresponding 
algebraic properties follow from the associated reality matrix $ \mathcal R $. 
We have shown that the above semisymplectic geometry turns out to be 
a well-defined configuration, which physically corresponds to the moduli 
dependent interacting statistical system. Mathematically, the above 
geometry describes a non-trivial deformed manifold and contains the 
effects of the moduli space $ \mathcal M_\phi $ and non trivially 
reduces to the thermodynamic like geometry, at the attractor fixed 
point(s) of the given black hole.

Finally, the present examination signifies geometric and algebraic
structures for the purely nonperturbative, topological and purely
quantum aspects and opens up guidelines towards the microstates
counting of a black hole. Thus, such a perspective not only offers
the role of the outer product of topological string wave function, 
but also intends to analyze the full density matrix of underlying 
quantum configurations. From the perspective of a definite quantum theory, 
we anticipate that there exist a rich class of rational conformal field 
theory (rCFT), which map Sen entropy function of a given black hole, 
as a generalized attractor mechanism, to the corresponding elliptic 
curves \cite{Moore2}. Thus, for a given entropy function 
$ F(\overrightarrow{s}) $, it would be interesting to understand 
how a rCFT configuration generalizes to the elliptic curves, 
or should the elliptic curves be themselves generalized, in order to 
yield a consistent rCFT spectrum of the extremal black holes? Secondly, 
from the viewpoint of flux compactifications \cite{Ooguri}, such a 
proposition is expected to offer simulating platforms in order to 
understand algebraic geometric properties of $ \mathcal N= 2 $ 
black hole attractors and integer SUSY attractors. In the above
exploitations, the Galois groups and Falting height functions are 
hereby undermined to offer interesting perspectives for nonzero 
heterotic dilaton-axion fields, apart from some discreet identifications. 
It would be interesting to push further the algebraic geometric 
perspectives of the generalized attractors, CFT configurations
and Sen entropy function.
\section*{Acknowledgements}
Towards the realization of this research, 
I would like to thank Prof. V. Ravishankar, Prof. S. Bellucci, Prof. P. Jain, 
Prof. U. B. Tewari, Prof. R. K. Thareja, Prof M. K. Harbola, Prof. S. G. Dhande 
for their support, help and encouragement.
Towards the motivation of this work, 
I would like to thank Prof. A. Sen for useful discussions 
on entropy function and entropy of black holes, offered during 
String Schools and String Workshops. I would like to thank 
Prof. Sen for  encouragement, teaching the subject of black hole physics
and discussions during \textit{``Advanced String School"}, held at 
\textit{``Institute of Physics Bhubaneswar, India''} 
and \textit{``Spring School on Superstring Theory and Related Topics''} held at 
\textit{``The Abdus Salam, International Centre for Theoretical Physics Trieste, Italy''}. 
In the cronological order of discussions, I would like to thank Prof. J. de Boer, 
Prof. S. Minwalla, Prof. A. Dabholkar, Prof. R. Gopakumar, Prof. V. V. Sridhar, 
Prof. S. Mahapatra, Prof. S. Siwach, Prof. S. Banerjee, Prof. A. Prakash, Prof. B. Bhattacharjya;
Prof. M. P. Lombardo,  Prof. S. Krishnakumar; and Dr. V. Chandra, Dr. P. Kumar, Dr. R. Singh, 
Dr. Y. K. Srivastava, Dr. A. Haque, R. Kumar, N. Gupta, N. Amuthan, and C. Srivastava 
for their discussions and suggestions on the geometric and algebraic aspects; 
and Dr. Mohd. A. Bhat, Dr. A. K. Mishra, N. K. Mishra and H. Ramraj for their 
suggestions and modifications in the langauge part.
I would like to hereby acknowledge the CSIR, New Delhi, India for doctoral research fellowship under the 
\textit{``CSIR-SRF-9/92(343)/2004-EMR-I"} at \textit{``Indian Institute of Technology Kanpur, India"}, 
where I enjoyed stimulating hospitality during my doctoral research; \textit{``The Abdus Salam, International 
Centre for Theoretical Physics Trieste, Italy"} towards my visits for the \textit{``Spring Schools 
on Superstring Theory and Related Topics''}; and \textit{``INFN-Laboratori Nazionali di Frascati, 
Roma, Italy''} for a postdoctoral research fellowship.

\end{document}